\RequirePackage{fix-cm}
\documentclass[smallextended]{svjour3}       
\smartqed  
\usepackage{relsize}
\usepackage{graphicx}
\usepackage{enumitem}
\usepackage{latexsym}
\usepackage{numprint}
\usepackage{url}

\usepackage{todonotes}
\usepackage{lipsum}
\usepackage{times}
\usepackage{latexsym}
\usepackage{listings}
\usepackage{booktabs}
\usepackage{multirow}
\usepackage{amsmath}
\usepackage{hyperref}
\usepackage{cancel}
\usepackage{amsmath}
\usepackage{tikz}

\lstdefinelanguage
   [x64]{Assembler}     
   [x86masm]{Assembler} 
   {morekeywords={CDQE,CQO,CMPSQ,CMPXCHG16B,JRCXZ,LODSQ,MOVSXD, %
                  POPFQ,PUSHFQ,SCASQ,STOSQ,IRETQ,RDTSCP,SWAPGS, %
                  rax,rdx,rcx,rbx,rsi,rdi,rsp,rbp, 
                  r8,r8d,r8w,r8b,r9,r9d,r9w,r9b, %
                  r10,r10d,r10w,r10b,r11,r11d,r11w,r11b, %
                  r12,r12d,r12w,r12b,r13,r13d,r13w,r13b, %
                  r14,r14d,r14w,r14b,r15,r15d,r15w,r15b,}} 

\usepackage{xcolor}
\usepackage{booktabs}
\usepackage{multirow,bigstrut}
\usepackage{tabu}

\definecolor{codegreen}{rgb}{0,0.6,0}
\definecolor{codegray}{rgb}{0.5,0.5,0.5}
\definecolor{codepurple}{rgb}{0.58,0,0.82}
\definecolor{backcolour}{rgb}{0.95,0.95,0.92}

\lstdefinestyle{mystyle}{
    backgroundcolor=\color{backcolour},   
    commentstyle=\color{codegreen},
    keywordstyle=\color{blue},
    numberstyle=\tiny\color{codegray},
    stringstyle=\color{codepurple},
    basicstyle=\ttfamily\footnotesize,
    breakatwhitespace=false,         
    breaklines=true,                 
    captionpos=b,                    
    keepspaces=true,                 
    numbers=left,                    
    numbersep=5pt,                  
    showspaces=false,                
    showstringspaces=false,
    showtabs=false,                  
    tabsize=2,
    columns=fullflexible,
}

\usepackage{comment}
\usepackage{xspace}
\newcommand{\datasetname}[1]{\emph{Shellcode\_IA32}}

\usepackage{amssymb}
\usepackage{amsmath}
\newcommand{\argmax}{\mathop{\mathrm{argmax}}}
\usepackage{soul,color}

\usepackage[caption=true, font=footnotesize]{subfig}
\def\BibTeX{{\rm B\kern-.05em{\sc i\kern-.025em b}\kern-.08em
    T\kern-.1667em\lower.7ex\hbox{E}\kern-.125emX}}

\usepackage{todonotes}
\usepackage{pdfcomment}

\begin{document}

\title{Can We Generate Shellcodes via Natural Language?\\An Empirical Study}


\author{Pietro Liguori         \and
        Erfan Al-Hossami \and
        Domenico Cotroneo \and
        Roberto Natella \and
        Bojan Cukic \and
        Samira Shaikh
}


\institute{P. Liguori, D. Cotroneo, R. Natella \at
              University of Naples Federico II, Naples, Italy \\
              \email{\{pietro.liguori, cotroneo, roberto.natella\}@unina.it}           
           \and
           E. Al-Hossami, B. Cukic, S. Shaikh  \at
              University of North Carolina at Charlotte, Charlotte, NC\\
              \email{\{ealhossa, bcukic, samirashaikh\}@uncc.edu}
}

\date{Received: date / Accepted: date}

\maketitle

\begin{abstract}
Writing software exploits is an important practice for \emph{offensive security} analysts to investigate and prevent attacks. 
In particular, \emph{shellcodes} are especially time-consuming and a technical challenge, as they are written in assembly language. 
In this work, we address the task of automatically generating shellcodes, starting purely from descriptions in natural language, by proposing an approach based on Neural Machine Translation (NMT). 
We then present an empirical study using a novel dataset (\datasetname{}), which consists of $3,200$ assembly code snippets of real Linux/x86 shellcodes from public databases, annotated using natural language. Moreover, we propose novel metrics to evaluate the accuracy of NMT at generating shellcodes. 
The empirical analysis shows that NMT can generate assembly code snippets from the natural language with high accuracy and that in many cases can generate entire shellcodes with no errors.

\end{abstract}

\section{Introduction}
\label{sec:introduction}
Nowadays, software security plays a crucial role in our society. Software vendors and users are in an arms race against cybercriminals, investing significant efforts towards identifying vulnerabilities and patching them, sometimes releasing updates mere hours after a release. 
The exploitation of software vulnerabilities is today a common \emph{offensive security} practice for security analysts, to understand how attackers take advantage of vulnerabilities, and to motivate vendors and users to patch them \cite{arce2004shellcode,mcgraw2004software,bugcrowd,hackerone}.
For example, in June 2021, GitHub updated its policy on malware and exploit research by allowing and even encouraging users to post \textit{proof-of-concept} (PoC) exploits or vulnerabilities on the platform~\cite{github_blog}.

Among software exploits, \textit{code-injection} attacks are considered the most dangerous ones, since they have the worst consequences on the victim organizations \cite{mason2009english}. 
Moreover, code-injection attacks have been drastically increasing with the growth of applications exposed to the Internet \cite{ray2012defining}, as shown by statistics from the Common Vulnerabilities and Exposures (CVE) database~\cite{cvedetails}. 
These attacks deliver and run malicious code (\textit{payload}) on the victims' machine, in order to give attackers control of the target system. Since the payload is typically designed to launch a command shell, the hacking community generically refers to the payload portion of a code-injection attack as a \textit{\textbf{shellcode}}.
Other objectives of shellcodes include killing or restarting other processes, causing a denial-of-service (e.g., a fork bomb), leaking secret data, etc. 
Listing~\ref{list:shellcode2} shows an example of shellcode\footnote{Shellcode collected from \url{https://www.exploit-db.com/shellcodes/48703}} in assembly for Linux OS running on the 32-bit Intel Architecture).

The development of software exploits is a technically difficult activity. Shellcodes are typically written in assembly language, in order to gain full control on the layout of code and data in stack and heap memory, to make the shellcode more compact, to obfuscate the code, and to perform low-level operations on data representation~\cite{deckard2005buffer,foster2005sockets,anley2007shellcoder,megahed2018penetration}. However, programming in assembly is time-consuming and has low productivity compared to high-level languages~\cite{dandamudi2005guide,jamwal2014c,pyeatt2016modern}. 

In order to make assembly programming easier and more efficient,  we investigate the use of Neural Machine Translation (NMT) for the generation of shellcodes. In general, NMT translates between different languages (including natural and programming languages), using Natural Language Processing (NLP) and Deep Learning (DL) techniques \cite{goodfellow2016deep,bahdanau2014neural,wu2016google,bojar2016findings}, in order to learn the typical idioms of a target programming language from datasets of annotated programs. 
NMT is an emerging approach for \emph{code generation} \cite{DBLP:journals/corr/YinN17,DBLP:journals/corr/LingGHKSWB16} and other programming tasks, such as code completion \cite{drosos2020wrex,shi_tf-coder_2020}, the generation of UNIX commands \cite{lin2017program,lin2018nl2bash} or commit messages \cite{jiang2017automatically,liu2018neural,jung2021commitbert}, etc.
However, NMT techniques have not heretofore been applied in the field of software security to generate software exploits.
In our case, developers would translate a description (\textit{intent}) of a piece of code in English, into the corresponding \textit{code snippet} in assembly language. 
For example, developers can use NMT to generate code snippets that they could not recall, or that are not yet confident to write themselves, similarly to querying a search engine, with the additional benefit of tailoring the code according to their query.

\begin{figure}[ht]
\begin{minipage}{\linewidth}
\small
\begin{lstlisting}[caption={Assembly code used to generate a shellcode on Linux OS running on 32 bit Intel Architecture. Lines 5-6, 11-12, 15-16, 19-20, 21-22-23, 24-25, 27-28-29 are multi-line snippets generated by seven different intents.},label={list:shellcode2}]
global _start;  	    Declare global _start.
section .text;  	    Declare code section.
_start:;        	    Define the _start label.
cld;            	    Clear the direction flag.
xor ecx, ecx;   	    Zero out the EAX register
mul ecx;              and the ECX register.	
incpage:;       	    Declare incpage function.
or cx, 0xfff;   	    Perform logical or between the CX register                          and 0xfff.
IncAddr:;       	    Declare the IncAddr label.
inc ecx;        	    Increment ECX.
push byte 0x43; 	    Put the syscall 0x43 into the EAX register.
pop eax	
int 0x80;       	    Execute execve syscall.
cmp al, 0xf2;         Jump to the IncPage label if the contents
jz IncPage;           of the AL register is equal to the value 0xf2.
mov eax, 0x50905090;	Move 0x50905090 into EAX.
mov edi, ecx;       	Move ECX into EDI.
scasd;                Jump to the IncAddr label if the value in the 
jnz IncAddr;          EAX register is not equal to the doubleword                         addressed by EDI
scasd;	              Jump to the IncAddr label if the value in
jnz IncAddr;          the EAX register is not equal to the doubleword                     addressed by EDI
jmp edi;              else jump to the EDI register.

xor ecx, ecx;       	Zero out the EAX register and the ECX register
mul ecx	
push eax;           	Push EAX on the stack.
push 0x68732f2f;    	Move ASCII /bin/sh into EBX.
push 0x6e69622f
mov ebx, esp	
mov al, 0xb;        	Move 0xb into AL.
int 0x80;           	call kernel

mov al, 0x01;       	Move 0x01 into AL.
xor ebx, ebx;       	Clear EBX.
int 0x80;           	Call kernel.
\end{lstlisting}
\end{minipage}
\end{figure}

In this paper, we introduce a novel approach for generating shellcodes in assembly language, from their description in natural language.
Differing from previous research, which adopts static and/or dynamic program analysis (e.g., fuzzing, program synthesis, etc.), we adopt a novel statistical, data-driven approach.  Specifically, our approach leverages state-of-the-art NMT techniques. Since NMT has never been applied to low-level languages such as assembly, our approach extends NMT by introducing an Intent Parser specialized for the assembly language and adopts transfer learning to bootstrap an NMT model from a training set of shellcodes. 
Then, the paper presents an extensive evaluation of the NMT approach. As there is no unique metric able to comprehensively represent the quality of translations, we introduce new metrics for this purpose. Indeed, the generated assembly code can have high accuracy compared to the ground truth, yet it may not be a working shellcode. Or, the generated program can be compilable and executable, but it may not implement the intended shellcode. Or again, the generated program does not exactly match the ground truth, but it can still be a correct shellcode (e.g., by using alternate valid labels or addressing modes), and so on. Therefore, we evaluate NMT from several points of view.

In summary, this work provides the following key contributions: 
\begin{itemize}[nolistsep,noitemsep]
    \item We propose a novel approach for translating natural language into shellcode in assembly language, based on NMT. The approach improves the state-of-the-art by using a novel, specialized Intent Parser and transfer learning. To the best of our knowledge, this is the first effort towards applying NMT to automatically generate code for security purposes;
    \item We release a curated, substantive corpus of real shellcodes from public databases, in order to support the training and evaluation of NMT systems for shellcode generation;
    \item We propose novel metrics to evaluate the performance of NMT systems for shellcode generation. Different from the metrics commonly used in other code generation tasks, the metrics proposed in this work go beyond evaluating performance on single-line snippets of code and also encompass the ability to generate entire, compilable shellcodes. Moreover, we look at the semantic correctness of the generated shellcode;
    \item We present an extensive empirical analysis of NMT techniques at generating shellcodes, supported by the proposed metrics and dataset.
\end{itemize}

In the following, Section~\ref{sec:related} discusses related work;
Section~\ref{sec:background} introduces background concepts;
Section~\ref{sec:approach} presents the proposed approach;
Section~\ref{sec:dataset} describes the dataset;
Sections~\ref{sec:experiments} experimentally evaluates the approach;
Section~\ref{sec:ethics} describes the ethical considerations;
Section~\ref{sec:threats} discusses the threats to validity of the work;
Section~\ref{sec:conclusion} concludes the paper.

\section{Related Work}
\label{sec:related}

Our work is situated at the intersection of machine translation and code/exploit generation, by applying NLP techniques to software security. Accordingly, we review related work in these areas. 

\vspace{0.1cm}
\noindent
\textbf{Neural Machine Translation for Code Generation} There are several recent works that focus on generating code from natural language \cite{yin2019reranking,dong2018coarse,DBLP:journals/corr/RabinovichSK17}. 
Ling \emph{et al.} \cite{DBLP:journals/corr/LingGHKSWB16} and Yin and Neubig \cite{DBLP:journals/corr/YinN17} proposed a novel neural architecture for code generation, while Xu \emph{et al.} \cite{Xu2020IncorporatingEK} incorporated pre-training and fine-tuning of a model to generate Python snippets from natural language using the CoNaLa dataset \cite{yin2018mining}.  Furthermore, Gemmell \emph{et al.}~\cite{gemmell2020relevance} used a transformer architecture with relevance feedback for code generation, and reported improvements over state-of-the-art on several datasets.
\\
There also exist approaches that perform the reverse task, i.e., generating natural language from code. 
Oda \emph{et al.} \cite{oda2015learning} pioneered the task of translating python code to pseudo-code while others proposed an n-gram language model to generate comments from source code \cite{movshovitz2013natural}. 
Iyer \emph{et al.} \cite{iyer2016summarizing} proposed an attention model that summarizes code.  Code2Seq~\cite{alon2018code2seq} embeds abstract syntax tree paths to encode context and was used for code documentation generation (generating natural language from code) and code summarization. 
A notable example of applying code documentation generation in software engineering is generating git commit messages from git-tracked codebase changes~\cite{jiang2017automatically}.
\\
NMT has been widely adopted also for different programming tasks. 
For example, Lin \emph{et al.} \cite{lin2018nl2bash} presented new data and semantic parsing methods to address the problem of mapping English sentences to \texttt{bash} commands, and Zhong \emph{et al.} \cite{Zhong2017Seq2SQLGS} generated SQL queries from natural language.
Tufano \textit{et al.}~\cite{tufano2019learning} investigated the ability of the NMT to learn how to automatically apply code changes implemented by developers during pull requests. 
The authors trained the model on a dataset containing pairs of code components before and after the implementation of the changes provided in the pull requests and showed that the NMT can accurately replicate the changes implemented by developers.
Hata \textit{et al.}~\cite{hata2018learning} presented Ratchet, an NMT-based technique that generates a fixed code for a given bug-prone code query. The technique uses a Seq2Seq model trained on pre-correction and post-correction code in past fixes. To prove the feasibility of the approach, the authors performed an empirical study on five open source projects, showing that Ratchet can generate syntactically valid statements with high accuracy.

Our empirical analysis investigates these recent advances in NMT in the context of the open problem of generating shellcodes in assembly language, from natural language intents. 




\vspace{0.1cm}
\noindent
\textbf{Automated Exploit Generation}.
The task of exploit generation via automatic techniques has been addressed in several ways. 
\textit{ShellSwap} \cite{bao2017your} is a system that generates new exploits based on existing ones, by modifying the original shellcode with arbitrary replacement shellcode. Hu \emph{et al.} \cite{hu2015automatic} developed a novel approach to construct data-oriented exploits through data flow stitching, by composing the benign data flows in an application via a memory error. They built a prototype attack generation tool that operates directly on Windows and Linux x86 binaries. 
Avgerinos et al. \cite{avgerinos2011aeg} developed an end-to-end system for automatic exploit generation (AEG) on real programs by exploring execution paths.  Given the potentially buggy program in source form, their proposal automatically looks for bugs, determines whether the bug is exploitable, and produces a working control-flow hijack exploit string. 
\textit{SemFuzz} \cite{you2017semfuzz} extracts necessary information from non-code text related to a vulnerability, using natural language processing and a semantics-based fuzzing process, in order to discover and trigger deep bugs. 
Chen \textit{et al.} \cite{chen2011automatic} presented techniques to find out the \textit{gadgets}, i.e., the basic building block in Jump Oriented Programming (JOP), and showed these gadgets are Turing complete. They implemented an automatic tool able to generate JOP shellcodes.
Ding et al. \cite{6999408} proposed a reverse derivation of a transformation method driven by state machines indicating the status of data flows, in order to transform the original shellcode into printable Return Oriented Programming (ROP) payload. 
\textit{Chainsaw} \cite{alhuzali2016chainsaw} is a tool for analyzing web applications and generating injection exploits. The tool performs static analysis and defines a model of the application behavior to generate injection exploits, by leveraging application workflow structures and database schemes.
Brumley et al. \cite{4531150} proposed an approach for Automatic Patch-based Exploit Generation (APEG). Starting from a program and its patched version, the approach identifies the security checks added by the patch and automatically generates inputs to fail the checks.
Huang et al. \cite{6717039} introduced a method to automatically generate exploits based on software crash analysis. This method analyzes software crashes using a symbolic failure model, to generate exploits from crash inputs and existing exploits for several types of applications.
Xu et al. \cite{8432013} developed a tool to find buffer overflow vulnerabilities in binary programs and automatically generate exploits using a constraint solver. Vulnerability detection is achieved through symbolic execution and the exploit generated by this tool can bypass different types of protection.

Similar to our previous work~\cite{liguori2021evil}, our approach uses natural language statements to generate exploits and adopts neither a static nor dynamic program analysis approach (e.g., fuzzing, program synthesis, etc.), but a statistical, data-driven approach.

\section{Background}
\label{sec:background}

This section introduces background concepts on neural machine translation (NMT). We follow the notation defined by Eisenstein \cite{eisenstein2018natural}.

\textit{Machine translation} refers to the translation of a language into another by the means of a computerized system \cite{dorr1999survey}.
It is defined as an optimization problem, which maximizes the conditional probability that a sentence $\omega^{(t)}$ in the target language is the likely translation of a sentence $\omega^{(s)}$ in the source language, by using a scoring function $\psi$:
\begin{equation}
    \hat{\omega}^{(t)} = \argmax_{\omega^{(t)}} \psi (\omega^{(s)}, \omega^{(t)}) 
\end{equation}
The resolution of the problem requires a decoding algorithm for computing $\hat{\omega}^{(t)}$, and a learning algorithm for estimating the parameters of the scoring function $\psi$.

\textit{Neural network models} for machine translation are based on the encoder-decoder architecture \cite{cho2014learning}. The encoder network converts the source language sentence into a context vector or matrix representation $z$ of fixed length. The decoder network then converts the encoding into a sentence in the target language by defining the conditional probability $p(\omega^{(t)}|\omega^{(s)})$.

The decoder is typically a recurrent neural network, which generates the target language sentence one word at a time, while recurrently updating a hidden state. The encoder and decoder networks are trained end-to-end from parallel sentences. If the output layer of the decoder is a logistic function, then the entire architecture can be trained to maximize the conditional log-likelihood:
\begin{gather}
    \log p(\omega^{(t)}| \omega^{(s)}) = \sum_{m=1}^{M^{(t)}} p(\omega_{m}^{(t)}| \omega_{1:m-1}^{(t)},z)\\
    p(\omega_{m}^{(t)}| \omega_{1:m-1}^{(t)},\omega^{(s)}) \propto \exp (\beta_{\omega_m^{(t)}}\cdot h_{m-1}^{(t)})
\end{gather}
where the hidden state $h_{m-1}^{(t)}$ is a recurrent function of the previously generated text $\omega_{1:m-1}^{(t)}$ and the encoding $z$, while $\beta \in R^{({V^{(t)\times K}})}$ is the matrix of output word vectors for the $V^{(t)}$ words in the target language vocabulary, and $K$ is the dimension of the hidden state.

\vspace{0.1cm}
\noindent
\textbf{Seq2Seq.}
The simplest encoder-decoder architecture is the sequence-to-sequence model \cite{sutskever2014sequence}. In this model, the encoder is set to the final hidden state of a long short-term memory (LSTM) \cite{hochreiter1997long} on the source sentence:
 \begin{gather}
     h_m^{(s)} = LSTM(x_m^{(s)},h_{m-1}^{(s)})\\
     z \triangleq h_{M^{(s)}}^{(s)} 
 \end{gather}
where $x_m^{(s)}$ is the embedding\footnote{The name is due to the fact that each word is embedded in a continuous vector space.} of the target language word $\omega_m^{(s)}$.
The encoding then provides the initial hidden state for the decoder LSTM:
\begin{gather}
    h_0^{(t)} = z \\
    h_m^{(t)} = LSTM (x_m^{(t)},h_{m-1}^{(t)})
\end{gather}
where $x_m^{(t)}$ is the embedding of the target language word $\omega_m^{(t)}$. 
Sequence-to-Sequence translation is nothing more than wiring together two LSTMs: one to read the source, and another to generate the target.

\vspace{0.1cm}
\noindent
\textbf{Attention Mechanism.} 
The weakness of using a fixed-length context vector is the difficulty to remember long sentences. Indeed, in the traditional Seq2Seq model, the intermediate states of the encoder are discarded, and only the final states (vector) are used to initialize the decoder.
To overcome this limitation, Bahdanau \textit{et al.} \cite{bahdanau2014neural} proposed the \textit{attention mechanism}, i.e., a solution that uses a context vector to align the source sentence and target sentence. The context vector holds the information from all hidden states from the encoder and aligns them with the current target output. By using this mechanism, the model is able to look at a specific part of the source sentence and better understand the relationship between the source and target.
\\
An \textit{attention function} can be described as mapping a query and a set of key-value pairs to an output, where the query, keys, values, and output are all vectors. The key-value-query concepts come from retrieval systems. For example, when a user types a query to search for a resource (value) on a contents-sharing platform, the search engine maps the query against a set of keys associated with the resources in the database of the platform and will show to the user the best-matched resource.  
Formally speaking, for each key $n$, the attention mechanism assigns a score $\sigma_a(m,n)$ with respect to the query $m$, based on how much they match. In Bahdanau’s paper, the score is parametrized by a feed-forward network with a single hidden layer.
The output of this activation function is a vector of non-negative numbers $[\alpha_{m\rightarrow 1}, \alpha_{m\rightarrow 2}, \dotsc , \alpha_{m\rightarrow N}]^T$, with length $N$ equal to the size of the memory (i.e., the space of all the generated words). 
Each value in the memory $v_n$ is multiplied by the attention $\alpha_{m\rightarrow n}$; the sum of these scaled values is the output. 
At each step m in decoding, the attentional state is computed by executing a query, which is equal to the state of the decoder, $h_m^{(t)}$. The resulting compatibility scores are:
\begin{gather}
    \psi_\alpha(m,n)=v_\alpha \cdot tanh(\Theta_\alpha|h_m^{(t)};h_n^{(s)})
\end{gather}

\vspace{0.1cm}
\noindent
\textbf{Transformer.}
In the encoder-decoder model, the keys and values used in the attention mechanism are the hidden state representations in the encoder network $z$, and the queries are state representations in the decoder network $h^{(t)}$.
Vaswani \textit{et al.} \cite{vaswani2017attention} proposed a new model architecture, the \textit{Transformer}, that does not rely on the recurrent neural networks by applying \textit{self-attention} \cite{lin2017structured,kim2017structured}) within the encoder and decoder. For level $i$, the basic equations of the encoder side of the transformer are:
\begin{gather}
    z_m^{(i)} = \sum_{n=1}^{M^{(s)}} \alpha_{m \rightarrow n}^{(i)}(\Theta_v h_n^{(i-1)})\\
    h_m^{(i)} = \Theta_2 ReLU(\Theta_1 z_m^{(i)} + b_1) + b_2
\end{gather}
For each token $m$ at level $i$, we compute self-attention over the entire source sentence.
The keys, values, and queries are all projections of the vector $h^{(i-1)}$.
The attention scores $\alpha_{m \rightarrow n}^{(i)}$ are computed using a scaled form of softmax attention.
This encourages the attention to be more evenly dispersed across the input. Self-attention is applied across multiple “heads”, each using different projections of $h^{(i-1)}$ to form the keys, values, and queries. 
The output of the self-attentional layer is the representation $z_m^{(i)}$, which is then passed through a two-layer feed-forward network, yielding the input to the next layer $h^{(i)}$.

The Transformer architecture first refines the input embedding of each token, by combining it with a \emph{positional encoding} vector. The architecture has a different positional encoding vector for each position of the sentence, in order to enrich the input embedding with positional information. Then, the transformed input embeddings sequentially go through the stacked encoder layers, which all apply a \emph{self-attention} process. The self-attention further refines an input embedding, by combining it with the other input embeddings for the sentence in a weighted way, in order to account for correlations among the words (e.g., to get information for a pronoun from the noun it refers to, the input embedding of the noun is given a large weight).

For more detailed information on NMT models, we refer the reader to the work of Eisenstein \cite{eisenstein2018natural}.

\section{Approach}
\label{sec:approach}
We leverage neural machine translation (NMT) to automatically generate shellcodes starting from their natural language description. 
Following prior work (e.g.,~\cite{luong2015effective}), we build a neural network that directly models the conditional probability of translating an \emph{intent}, in natural language into a \emph{code snippet} in assembly language.

The main challenge towards the goal of automatically generating shellcodes is represented by the programming language, i.e., the assembly. This language is significantly different from other languages addressed so far by research on NMT, which focused so far on mainstream imperative languages such as Python and Java. Assembly is a low-level programming language with many syntactical differences from these languages. For example, assembly does not provide the concept of variable, which is instead replaced by registers, memory addresses, addressing modes, and labels. Moreover, some programming constructs in assembly require multiple statements, which instead could be expressed with only one statement of other programming languages.
To address this new language for NMT, we opted to base our solution on existing deep neural network architectures: \textit{Seq2Seq with Attention}, and \textit{CodeBERT}. 
\\
We refrained from proposing a new architecture, for several reasons: (i) using an existing, well-tested architecture can be used with more confidence in a comparative setting in which numerical issues (such as, the \textit{vanishing gradient}) can be prevented; (ii) existing architectures were shown to perform well when translating from English descriptions, which is also the case of our problem; (iii) using an existing architecture enables us to reuse pre-trained models, which are costly to pre-train from scratch in terms of data size, computational time, and resources.
\\
Furthermore, assembly is a low-resource programming language and its codebases are scarce data compared to mainstream program languages and, therefore, it would be a challenge to pre-train a model from scratch on assembly-based shellcode bases. Since NMT for assembly code-based shellcodes is not investigated in prior works, there are limited resources for processing assembly codebases such as abstract syntax trees (AST), which are abundant for other programming languages and provide domain knowledge for some existing code generation architectures. Due to these reasons, we hence wanted to thoroughly investigate the strengths and weaknesses of current architectures. In the following, we briefly describe these architectures.

\vspace{0.1cm}
\noindent
\textbf{Seq2Seq} is a common model used in a variety of neural machine translation tasks. Similar to the encoder-decoder architecture with Bahdanau's attention mechanism \cite{bahdanau2014neural}, we use a bi-directional LSTM as the encoder, to transform an embedded intent sequence into a vector of hidden states with equal length. 
Within the bidirectional LSTM encoder, each hidden state corresponds to an embedded token. The encoder LSTM is bidirectional, which means it reads the source sequence ordered from left to right and from right to left.
To combine both directions, each hidden state for the bidirectional LSTM encoder is computed by concatenating the forward and backward hidden states in the encoder.

\vspace{0.1cm}
\noindent
\textbf{CodeBERT}~\cite{feng_codebert_2020} is a large multi-layer bidirectional Transformer architecture~\cite{vaswani2017attention}. Like Seq2Seq, the Transformer architecture is made up of encoders and decoders. CodeBERT has 12 stacked encoders and 6 stacked decoders. Compared to Seq2Seq, the Transformer architecture introduces mechanisms to address key issues in machine translation: (i) the translation of a word depends on its position within the sentence; (ii) in the target language, the order of the words (e.g., adjectives before a noun) can be different from the order of words in the source language (e.g., adjectives after a noun); (iii) several words in the same sentence can be correlated (e.g., pronouns). These problems are especially important when dealing with long sentences.
Different from Seq2Seq, CodeBERT also comes with a \emph{pre-trained} neural network model, learned from large amounts of code snippets and their descriptions in the English language, and covering six different programming languages, including Python, Java, Javascript, Go, PHP, and Ruby. The goal of pre-training is to bootstrap the training process, by establishing an initial version of the neural network, to be further trained for the specific task of interest \cite{peters2018deep,liu2019roberta,devlin2018bert,brown2020language}. This approach is called \emph{transfer learning}. 
In our case, we train the CodeBERT model to translate English intents to assembly code snippets using our dataset (see \S{}~\ref{sec:dataset}).

\begin{figure*}[ht]
\centering
\includegraphics[width=1\linewidth]{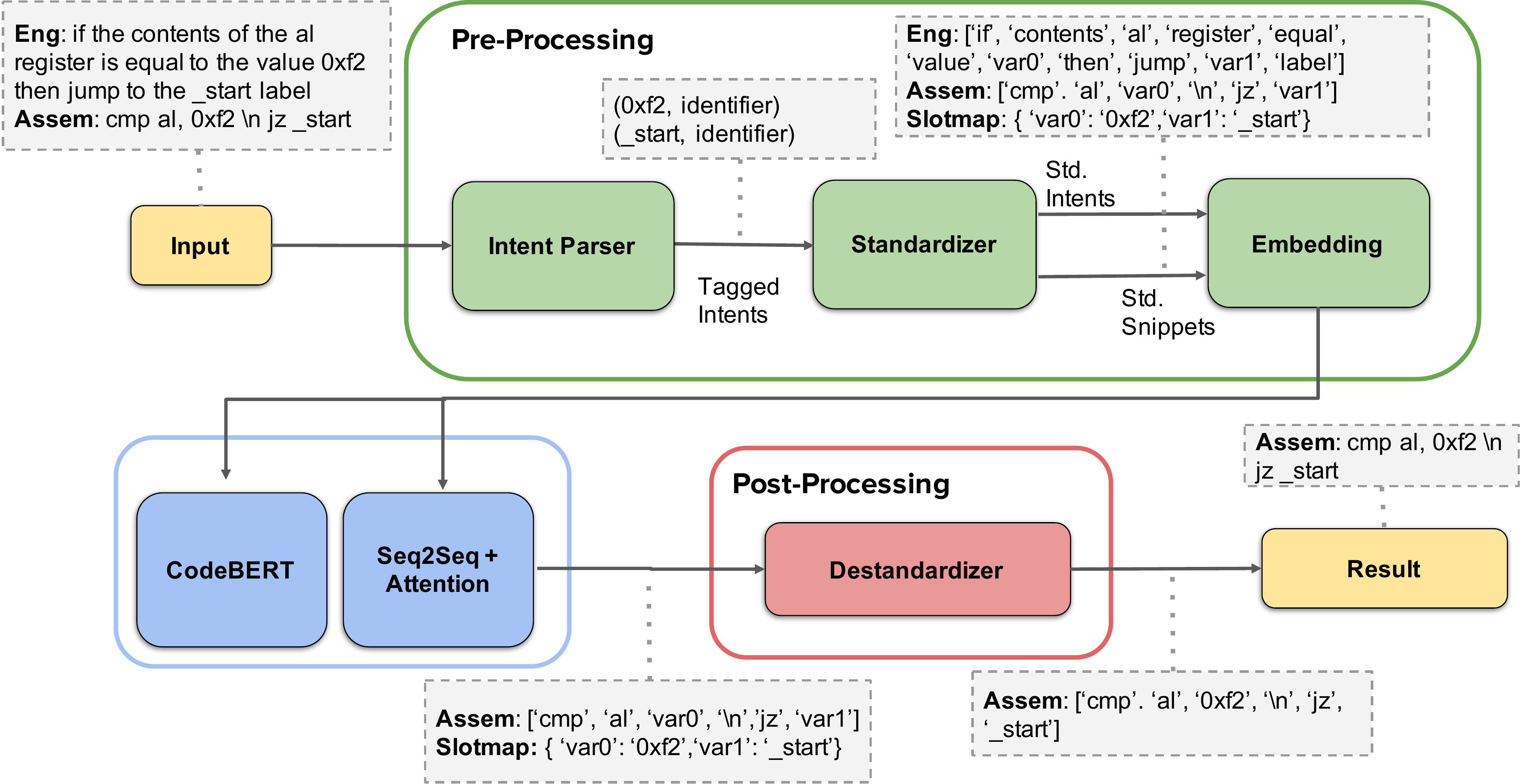}
\caption{Diagram showing the steps of the approach: 1) Pre-Processing of intent-code samples in both training and validation sets, 2) translation of unseen intent samples from the validation-set, and lastly, 3) Post-Processing applied to generated samples}
\label{fig:approach}
\end{figure*}

To better support such existing models at performing a new translation task, we extended the process with data processing. Data processing is an essential step to support the NMT models in the automatic code generation and refers to all the operations performed on the data used to train, validate and test the models.
These operations strongly depend on the specific source and target languages to translate (in our case, English and assembly language). We process data through a pipeline of steps, which we tailored for the task of generating assembly code snippets. 
The data processing steps are performed both before translation (\textit{pre-processing}), to train the NMT model and prepare the input data, and after translation (\textit{post-processing}), to improve the quality and the readability of the code in output.
Figure~\ref{fig:approach} shows the architecture of our approach, along with an example of inputs and outputs at each step, further discussed in the following.

\subsection{Pre-Processing}
\label{subsec:preprocessing}
The pre-processing starts with the \textit{stopwords filtering}, i.e., by removing a set of custom compiled words (e.g., \textit{the, each, onto}), in order to include only relevant data for machine translation. 
This phase also includes the identification of \textit{tokens}, i.e., basic units which need not be decomposed in subsequent processing. Therefore, the input sequences of natural language tokens and assembly code are split in a process called \textit{tokenization}. The tokenizer converts the input strings into their byte representations, and learns to break down a word into subword tokens (e.g., lower becomes \texttt{[low,er]}. 
We tokenize intents using the \textit{nltk word tokenizer} \cite{loper2002nltk} and snippets using the Python \textit{tokenize} package \cite{tokenize}.

One task for code generation systems is to prevent non-English tokens (e.g., \texttt{\_start}) from getting transformed during the learning process. This process is known as \textit{object standardization}. Abstracting important words for the assembly language can make it easier for the model to reuse existing structures learned from other imperative languages, such as moving data and changing the control flow.
To perform the standardization, we adopt an \textit{intent parser}, which takes in input a natural language intents and provides as output a dictionary of standardizable tokens (i.e., it identifies the correct names for the standardization process), such as the names of the registers, the actions (e.g., \texttt{/bin/sh}), the hexadecimal values, etc. 
We implement the intent parser using \emph{spaCy}, an open-source, industrial-strength Natural Language Processing library written in Python and Cython. 
We also use custom rules defined with regular expressions to identify hexadecimal values (e.g., \texttt{0xbb}), strings that fall between quotation marks, squared brackets, variable name notations (e.g., \texttt{variableName}, \texttt{variable\_name}), function and register names, mathematical expressions, and byte arrays (e.g., \texttt{\textbackslash xe3 \textbackslash xa1}). Hence, this component is tailored for the task of generating shellcodes in assembly language starting from their natural language description.

All tokens selected by the parser are therefore passed to the \textit{Standardizer}. The standardization process simply replaces the selected token in both the intent and snippet with \texttt{var\#}, with \texttt{\#} denoting a number from $0$ to $|l|$, and $|l|$ is the number of tokens to standardize.  
In \figurename~\ref{fig:approach}, the intent parser identifies \texttt{0xf2}, and \texttt{\_start} as standardizable tokens and standardizes them to \texttt{var0}, and \texttt{var1} respectively (based on order of appearance in the intent). 
To improve the process, we prevent the standardization of unimportant tokens, by compiling a dictionary of $45$ assembly keywords (e.g., \texttt{register}, \texttt{address}, \texttt{byte}, etc.) as non-standardizable tokens. 
After the standardization process, both the original token and its standardized counterpart (\texttt{var\#}) are stored in a dictionary (named \textit{Slotmap}) to be used during post-processing to restore the original words.

Lastly, we create \textit{word embeddings}, i.e., we map each token (in both the intent and code snippet sequences) into a numerical id representation in order to capture their semantic and syntactic information, where the semantic information correlates with the meaning of the tokens, while the syntactic one refers to their structural roles \cite{li2018word}.

\subsection{Post-Processing}
\label{subsec:postprocessing}
Post-processing is an automatic post-editing process, applied during decoding in the translation process (i.e., after the generation of the code snippet).
This phase include a \textit{Destandardizer}, which uses the slot map dictionary generated by the parser to replace all keys in the standardized intent (i.e., \texttt{var0} and \texttt{var1}) with the corresponding memorized values (i.e., \texttt{0xf2}, and \texttt{\_start}). 

The generated snippets are then further post-processed using regular expressions. This operation includes the removal of (any) extra-spaces in the output (e.g., between operations and operands), and the removal of (any) extra-backslashes in escaped characters (e.g., \texttt{\textbackslash{\textbackslash{n}}}). Also, during the post-processing, newline characters \texttt{\textbackslash{n}} are replaced with new lines to generate multi-line snippets. 
As a final step, snippet tokens are joined to form a complete code snippet.

\section{Dataset}
\label{sec:dataset}
We curated and released a dataset for, \textbf{\datasetname{}} \cite{liguori2021shellcode}, specific to shellcode generation. This dataset consists of $3,200$ examples of instructions in assembly language for \textit{IA-32} (the 32-bit version of the x86 Intel Architecture) collected from publicly available security exploits. The x86 is a complex instruction set computer (CISC), in which single instructions can perform several low-level operations (such as a load from memory, an arithmetic operation, and a memory store) or are capable of multi-step operations or addressing modes within single instructions. 
The dataset is comparable in size to the popular CoNaLa dataset~\cite{yin2017syntactic} ($2,379$ training and $500$ test samples in the \textit{annotated} version of the dataset), which is the basis for state-of-the-art studies in NMT for Python code generation \cite{yin2018mining,yin2019reranking,gemmell2020relevance}.

\begin{figure}[ht]
    \centering
    \includegraphics[width=0.75\columnwidth]{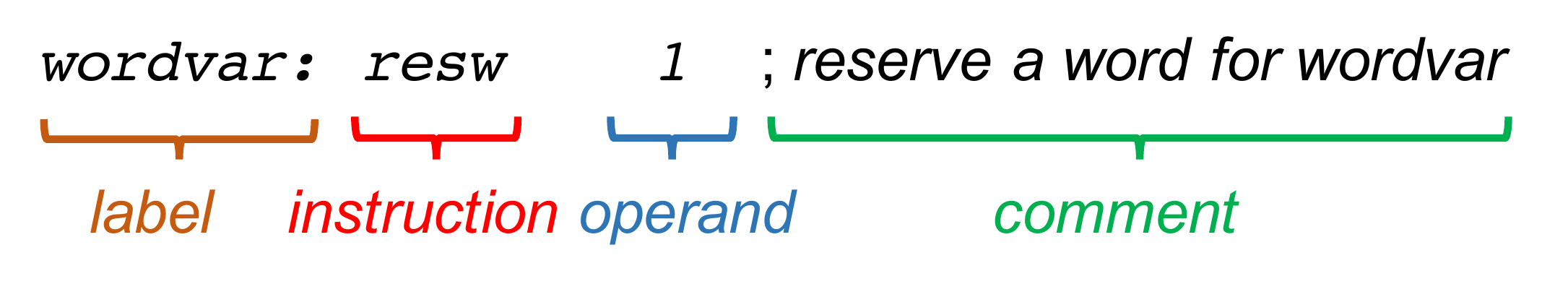}
    \caption{Layout of a NASM source line}
    \label{fig:assembly_instruction}
\end{figure}

We collected assembly programs used to generate shellcode from \textit{shell-storm} \cite{shellstorm} and from \textit{Exploit Database} \cite{exploitdb}, in the period between August 2000 and July 2020.
We focus on shellcode for Linux, the most common OS for security-critical network services. Accordingly, we gathered assembly instructions written for the \textit{Netwide Assembler} (NASM) for Linux \cite{duntemann2000assembly}.
NASM is a line-based assembler. Figure~\ref{fig:assembly_instruction} shows a simple example of a NASM source line. Every source line contains a combination of four fields: an optional \textit{label}, to symbolically represent the address of an opcode or data location defined by the line; a \textit{mnemonic} or \textit{instruction}, which identifies the purpose of the statement and is optionally followed by \textit{operands} specifying the data to be manipulated; an optional \textit{comment}, i.e., free text ignored by the compiler. A mnemonic is not required if a line contains only a label or a comment. 
 
The assembly programs collected in the dataset implement a varied set of shellcode attacks. One of the most common and basic shellcodes is the execution of a system shell (e.g., the \texttt{/bin/sh} command). This shellcode is often used in combination with more sophisticated attacks. The main categories include: exfiltrating password, e.g., from \texttt{/etc/passwd} (a plain text-based database that contains information for all user accounts on the system); breaking a chroot jail (an additional layer of security to run untrusted programs, which can be evaded by invoking vulnerable system calls with malicious inputs); running executables with the file system permissions of the executable's owner; flushing firewall rules (e.g., IPtables). Another form of shellcodes is the \textit{egg hunter}, i.e., a piece of code that when executed looks for other pieces of code (usually bigger) called the \textit{egg} and passes the execution to the egg. This technique is usually used when the space of executing shellcode is limited (the available space is less than the egg size) and it is possible to inject the egg into another memory location. Shellcodes are also used to perform \textit{denial-of-service} (DoS) attacks, such as for the \textit{fork-bomb} attack, in which a process continually replicates itself to deplete system resources, slowing down or crashing the system due to resource exhaustion. Among the most complex shellcodes, we find the \textit{bind shell} attacks. These attacks, which can easily reach hundreds of bytes, are used to open up a port on the victim system and connect to it from the remote attacking box. The complexity further increases when an attack redirects all inputs and outputs to a socket (\textit{reverse shell}) in order to evade firewalls.

Each sample of the \datasetname{} dataset represents a snippet~\textendash~intent pair. The \textbf{snippet} is a line or a combination of multiple lines of assembly code, following the NASM syntax. The \textbf{intent} is a comment in the English language (c.f. Listing~\ref{list:shellcode2}).
To take into account the variability of descriptions in natural language, multiple authors described independently different samples of the dataset in the English language. Where available, we used as natural language descriptions the comments written by developers of the collected programs. 
Moreover, in the preliminary phase of the dataset collection, we enriched the dataset with lines of assembly code and their relative English comments extracted from popular tutorials and books~\cite{duntemann2011assembly,kusswurm2014modern,tutorialspoint}. 
This helped us to learn the typical style for describing assembly code and to mitigate bias in our descriptions in English of assembly code. Once we reached confidence about the description style (i.e., the description style was recurring when adding more samples), we focused our efforts on real shellcodes, by writing ourselves the descriptions where no comment or documentation about the code snippet was available. Our dataset consists of $~10\%$ of instructions collected from books and guidelines, while the rest are from real shellcodes. However, there is no qualitative difference between both sets.

\vspace{0.1cm}
\noindent
\textbf{Multi-line Snippets:} Since assembly is a low-level language, it is often necessary to use multiple instructions to perform a given task. Thus, we go beyond one-to-one mappings between a line of code and its comment/intent. For example, a common operation in shellcodes is to save the ASCII string \textit{``/bin/sh"} into a register. This operation requires three distinct assembly instructions: push the hexadecimal values of the words \textit{``/bin"} and \textit{``//sh"} onto the stack register before moving the contents of the stack register into the destination register (lines 27-28-29 in Listing~\ref{list:shellcode2}). 
It would be meaningless to consider these three instructions as separate. To address such situations, we include $510$ lines ($\sim16\%$ of the dataset) of intents that generate multiple lines of shellcodes (separated by the newline character \textbackslash{n}). Table~\ref{tab:multiple_snippets} shows two further examples of multi-line snippets with their natural language intent.

\begin{table*}[ht]
\centering
\caption{Examples of multi-line snippets}
\label{tab:multiple_snippets}
\begin{tabular}
{>{\centering\arraybackslash}m{5.5cm} | 
>{\centering\arraybackslash}m{5.5cm}}
\toprule
\textbf{English Intent} & \textbf{Multi-line Snippets} \\ \midrule
\textit{jump short to the decode label if the contents of the al register is not equal to the contents of the cl register else jump to the shellcode label} & \texttt{cmp al, cl} \textbackslash{}n \texttt{jne short decode} \textbackslash{}n  \texttt{jmp shellcode} \\ \midrule
\textit{jump to the label recv\textunderscore http\textunderscore request if the contents of the eax register is not zero else subtract he value 0x6 from the contents of the ecx register} & 
\texttt{test eax, eax} \textbackslash{}n \texttt{jnz recv\textunderscore http\textunderscore request} \textbackslash{}n   \texttt{sub ecx, 0x6}\\ \bottomrule
\end{tabular}
\end{table*}

\begin{figure}[ht]
    \centering
    \subfloat[Number of tokens in the intents.]{%
    \includegraphics[width=0.5\columnwidth]{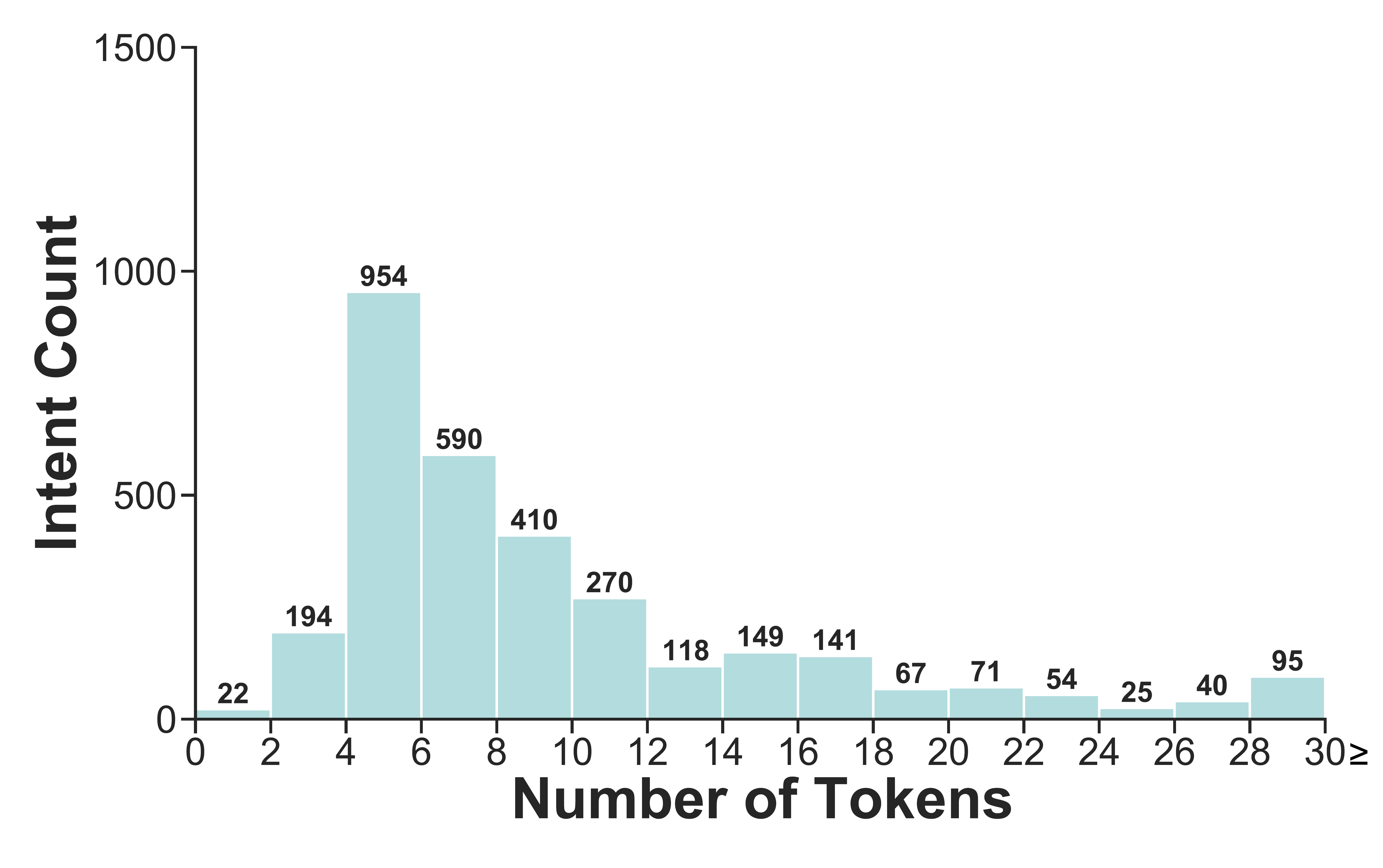}}
    \subfloat[Number of tokens in the snippets.]{%
    \includegraphics[width=0.5\columnwidth]{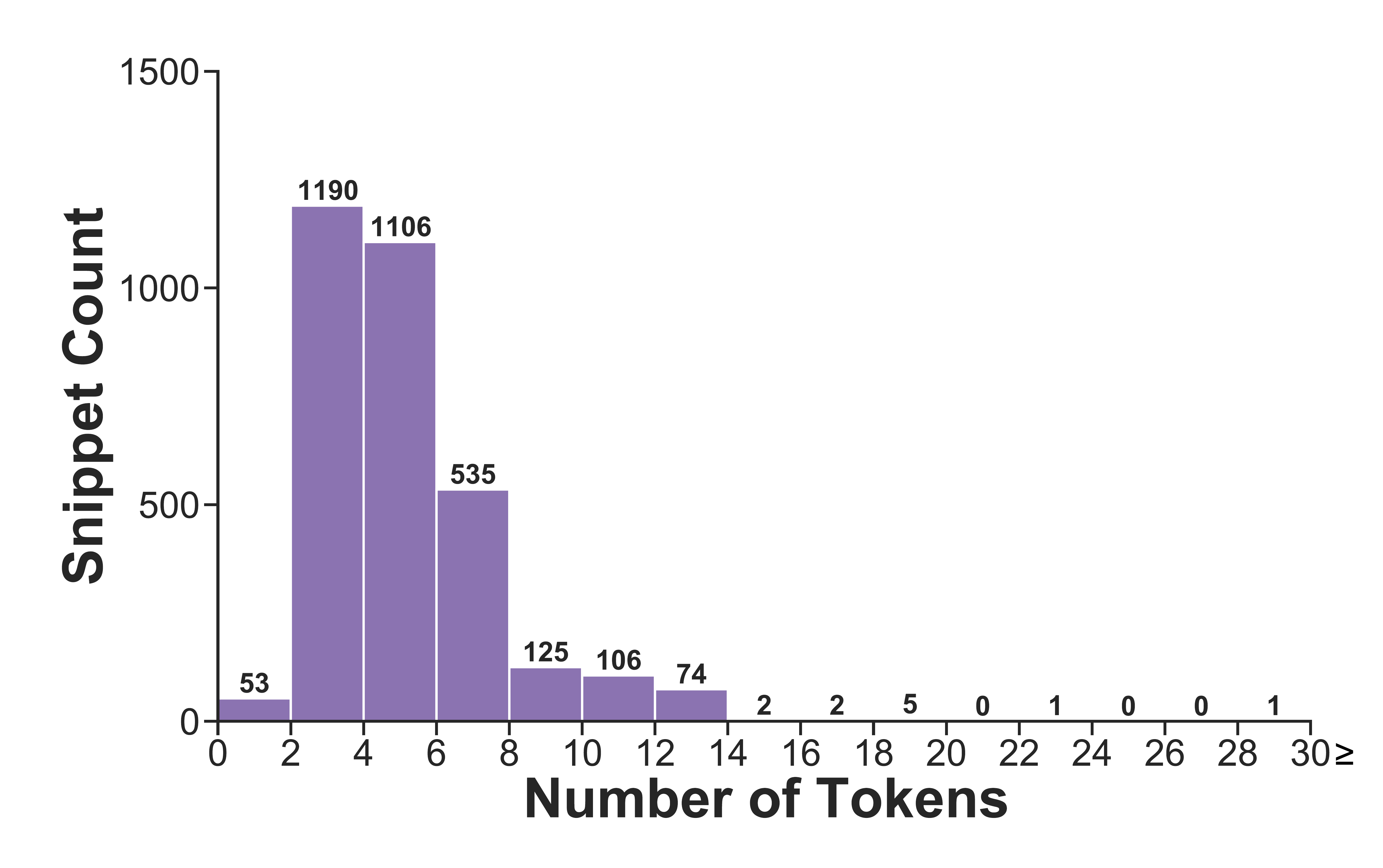}}\hfill
    \caption{Histogram of the \datasetname{} dataset showcasing the distribution of token counts across intents and snippets.}
    \label{fig:data_hist}
     
\end{figure}

\vspace{0.1cm}
\noindent
\noindent
\textbf{Statistics:} Table \ref{tab:dataset_statistics} presents  descriptive statistics of the \datasetname{} dataset.
The dataset contains $52$ distinct assembly mnemonics, excluding declarations of functions, sections, and labels.  
The two most frequent assembly instructions are \texttt{mov} ($\sim30$\% frequency), used to move data into/from registers/memory or to invoke a system call, and \texttt{push} ($\sim22$\% frequency), which is used to push a value onto the stack. The next most frequent instructions are the \texttt{cmp} ($\sim 7\%$ frequency), \texttt{xor} and \texttt{jmp} instructions ($\sim 4\%$ frequency). 
The \textit{low-frequency words} (i.e., the words that appear only once or twice in the dataset) contribute to the $3.6\%$ and $7.3\%$ of the natural language and the assembly language, respectively. 
\figurename{}~\ref{fig:data_hist} shows the distribution of the number of tokens across the intents and snippets in the dataset.
We publicly shared our entire \datasetname{} dataset on a GitHub repository.\footnote{The dataset can be found here: \url{https://github.com/dessertlab/Shellcode_IA32}}

\begin{table}[ht]
\centering
\caption{\datasetname{} statistics}
\label{tab:dataset_statistics}
\begin{tabular}{
 >{\centering\arraybackslash}m{1.5cm}
 >{\centering\arraybackslash}m{1.5cm}
 >{\centering\arraybackslash}m{1.5cm}
 >{\centering\arraybackslash}m{1.5cm}
 >{\centering\arraybackslash}m{1.5cm}
 >{\centering\arraybackslash}m{1.5cm}}
\toprule
\textbf{Language} & \textbf{Unique statements} & \textbf{Unique tokens} & \textbf{Avg. tokens per statement} & \textbf{Min tokens per statement} & \textbf{Max tokens per statement}\\
\midrule
\textit{Natural Language} & 3,184 & 1639 & 9.15 & 1 & 46\\
\midrule
\textit{Assembly Language} & 2,248 & 1401 & 4.17 & 2 & 30\\
\bottomrule
\end{tabular}
\end{table}

\section{Experimental Analysis}
\label{sec:experiments}
This section presents an extensive evaluation of our approach to generating shellcodes from natural language descriptions. 
We conducted the experimental analysis to target the following experimental objectives.

\vspace{0.1cm}
\noindent
$\rhd$ \textit{Feasibility in applying NMT for shellcode generation.}\\
We first perform an initial assessment on the feasibility of using NMT for shellcode generation with reasonably good accuracy, by applying techniques commonly used for code generation (e.g., generating Python code from natural language). 
We evaluate a broad set of state-of-the-art models for code generation, in combination with different techniques for data processing. In this initial stage, we adopt automatic evaluation metrics.

\vspace{0.1cm}
\noindent
$\rhd$ \textit{Accuracy of NMT at generating assembly code snippets.}\\
In this experimental objective, we deepen the analysis of the accuracy of NMT models. 
This is a cumbersome task since automatic metrics do not catch the deeper linguistic features of generated code, such as its semantic correctness \cite{DBLP:journals/corr/abs-2105-03311}. Therefore, it is also advisable for NMT studies to perform an evaluation through manual analysis, by using additional metrics in order to have a more precise and complete evaluation. 
The second experimental objective still focuses on the analysis of individual intents and their corresponding translations into code snippets. 


\vspace{0.1cm}
\noindent
$\rhd$ \textit{Accuracy of the NMT at generating whole shellcodes.}\\
We investigate if it is possible to apply NMT to generate full shellcodes, i.e., entire assembly programs from a set of intents. 
Ideally, the generated code is entirely or mostly correct, in order to reduce the human effort towards developing assembly programs. 
Therefore, in this experimental objective, we evaluate how many entire shellcodes are correctly generated by NMT (unlike the previous experimental objective, where we analyze individual code snippets regardless of which program they belong to).

\vspace{0.1cm}
\noindent
$\rhd$ \textit{Types of errors incurred by NMT in the generation of shellcodes.}\\
In this experimental objective, we are concerned with diagnosing the error predictions in the code generation task. We qualitatively analyze a representative sample of the most frequent mistakes, including both syntactic and semantic ones, to get more insight into the severity of the errors, and to understand potential areas of improvement for future work.

\subsection{Model Implementation}
We implement the Seq2Seq model using {\fontfamily{qcr}\selectfont xnmt} \cite{neubig2018xnmt}. We use an Adam optimizer \cite{kingma2014adam} with $\beta_1=0.9$ and $\beta_2=0.999$, while the learning rate $\alpha$ is set to $0.001$. We set all the remaining hyper-parameters in a basic configuration: layer dimension = $512$, layers = $1$, epochs (with early stopping enforced) = $200$, beam size = $5$. 

Our CodeBERT implementation uses an encoder-decoder framework where the encoder is initialized to the pre-trained CodeBERT weights, and the decoder is a transformer decoder. The decoder is composed of $ 6$ stacked layers. The encoder follows the RoBERTa architecture~\cite{liu2019roberta}, with $12$ attention heads, hidden layer dimension of $768$, $12$ encoder layers, $514$ for the size of position embeddings. We use the Adam optimizer~\cite{kingma2014adam}. The total number of parameters is 125M. The max length of the input is $256$ and the max length of inference is $128$. The learning rate $\alpha = 0.00005$, batch size = $32$, beam size = $10$, and train\_steps = $2800$.

We performed our experiments on a Linux machine. Seq2seq utilized $8$ CPU cores and $8$ GB RAM. CodeBERT utilized 8 CPU cores, $16$ GB RAM, and $2$ GTX1080Ti GPUs.
The computational time needed to generate the output depends on the settings of the hyper-parameters and the size of the dataset. 
On average, the training time for the Seq2Seq model was $\sim60$ minutes, while CodeBERT required for the training on average $\sim220$ minutes.
Once the models are trained, the time to translate intent into a code snippet is below 1 second and can be considered negligible. 

\subsection{Test Set}
To perform the experimental evaluation, we split our entire dataset into train/dev/test sets by using an $80$/$10$/$10$ ratio.
To divide the data between training, dev, and test set, we did not individually sample intent-snippet pairs from the dataset, but we took groups of intent-snippet pairs that belonged to the same shellcode, in order to be able to evaluate generate shellcodes in their entirety (see \S~\ref{subsec:RQ3}). The test set contains $30$ complete shellcodes (e.g. the entire Listing~\ref{list:shellcode2}). 

\begin{figure}[ht]
    \centering
    \subfloat[Number of assembly lines of code.\label{fig:program_length}]{%
    \includegraphics[width=0.5\columnwidth]{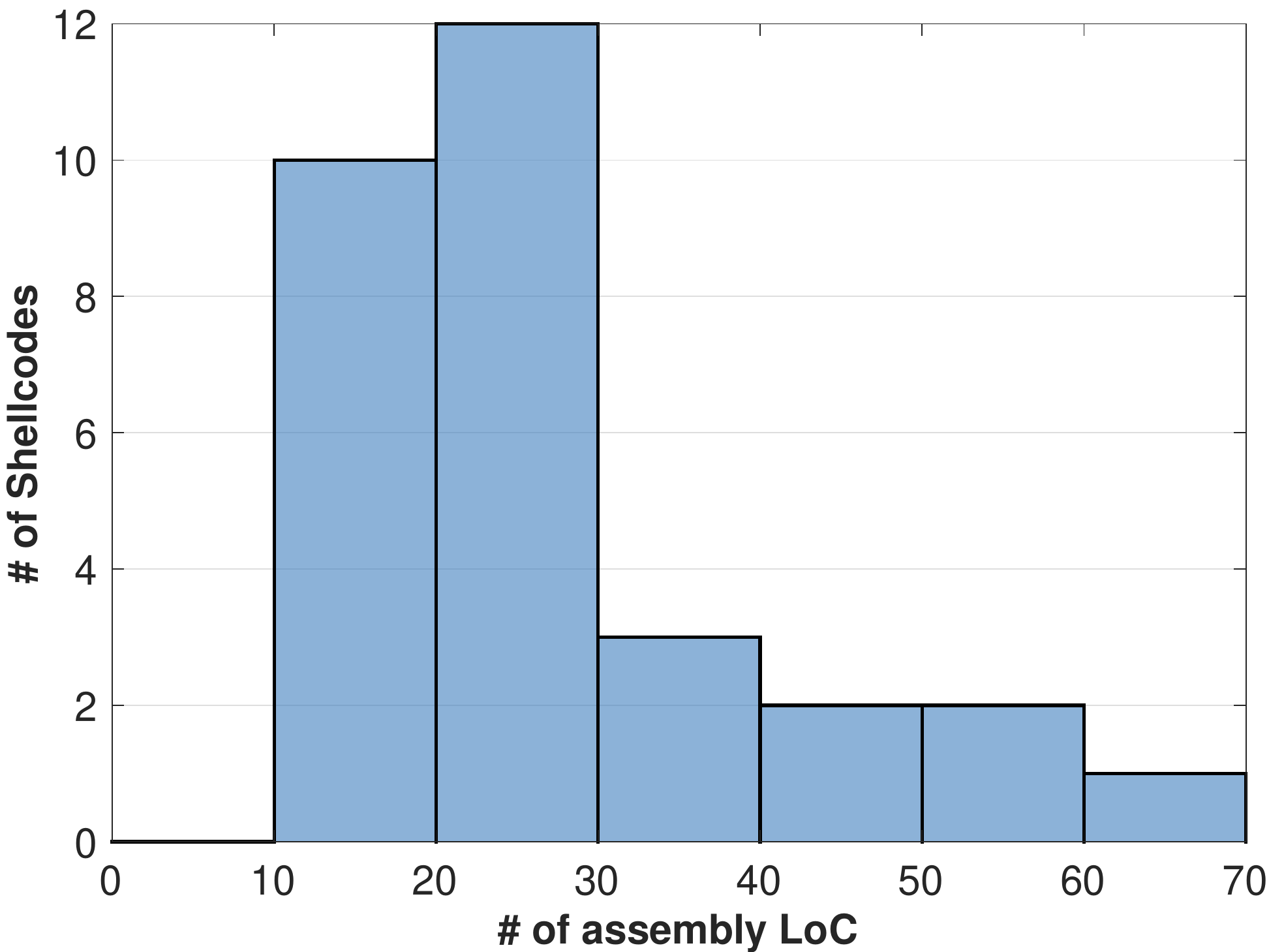}}
    \subfloat[Number of multi-lines snippets.\label{fig:multiline}]{%
    \includegraphics[width=0.5\columnwidth]{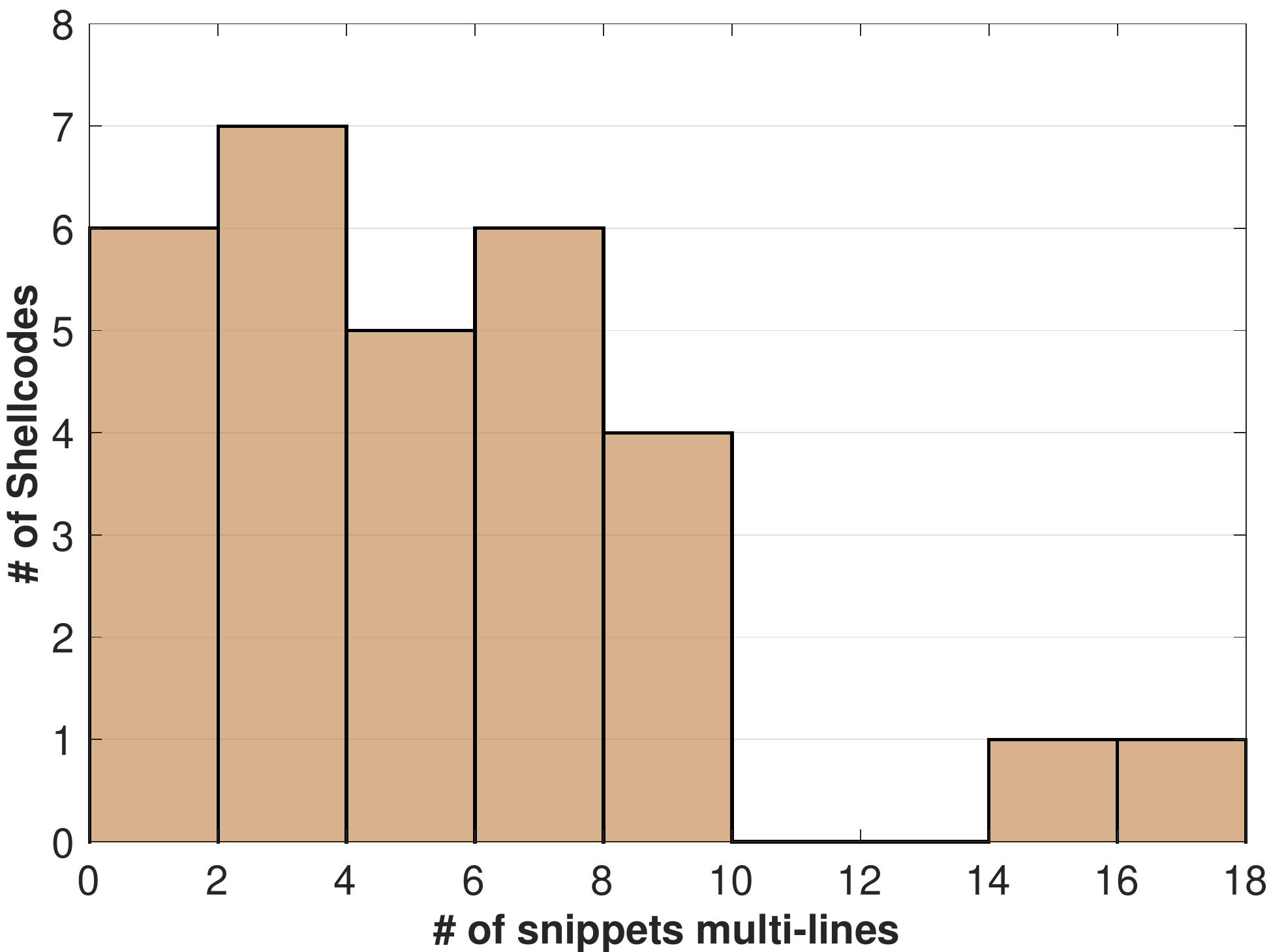}}\hfill
    \caption{Histograms visualizing the statistics of the  30 shellcodes in the test set.}
    \label{fig:test_statistic}
     
\end{figure}

We selected the 30 shellcodes of the test set in order to maximize the heterogeneity among the programs and mitigate bias. We anticipated that these biases could affect the evaluation: the type of attack (as they may entail different instructions and constructs); the authors of the shellcode (as it may also affect the programming style); and the complexity of the shellcode (as more complex shellcodes may also be more difficult to describe and to translate). We divided the shellcodes according to the type of the attack (shell spawning, break chroot, fork bomb, etc.), and sampled the shellcodes uniformly across these classes. When sampling within each class, we double-checked that no programmer was over-represented. We used the shellcode length as a proxy for complexity, and we increased the sample size until the distribution of the shellcode length was comparable to the distribution of the whole population (min=$12$, max=$61$, mean=$26.9$, median=$24.5$).
The histograms in \figurename{}~\ref{fig:test_statistic} summarize the statistic of the programs in the test set in terms of lines of code.
Additional information on the test set is presented in the \appendixname~\ref{appendix:test_set}. 

\subsection{Feasibility in applying NMT for shellcode generation}
\label{subsec:RQ1}

We first analyze the feasibility of Seq2Seq with attention mechanism and CodeBERT for the generation of shellcodes and investigate the impact of the data processing described in \S{}~\ref{sec:approach}. In this stage, we use automatic evaluation metrics. 
Automatic metrics are commonly used in the field of machine translation. They are reproducible, easy to be tuned, and time-saving. The \emph{BiLingual Evaluation Understudy} (BLEU) \cite{papineni2002bleu} score is one of the most popular automatic metric \cite{oda2015learning,DBLP:journals/corr/LingGHKSWB16,gemmell2020relevance,tran2019does}. This metric is based on the concept of \textit{n-gram}, i.e., the adjacent sequence of $n$ \textit{items} (e.g., syllables, letters, words, etc.) from a given example of text or speech. In particular, this metric measures the degree of n-gram overlapping between the strings of words produced by the model and the human translation references at the corpus level. BLEU measures translation quality by the accuracy of translating n-grams to n-grams, for n-gram of size $1$ to $4$ \cite{han2016machine}.
The \emph{Exact match accuracy} (ACC) is another automatic metric often used for evaluating neural machine translation  \cite{DBLP:journals/corr/LingGHKSWB16,yin2017syntactic,yin2018tranx,yin2019reranking}. It measures the fraction of the exact match between the output predicted by the model and the reference.

To assess the influence of our tailoring to NMT for the assembly language (e.g., the intent parser), we compare three ``variants'' of NMT by varying the steps of the data processing pipeline (see \S~\ref{sec:approach}):
\begin{itemize}
    \item \textit{w/o data processing}: the model performs the translation task without applying any step of the data processing pipeline. 
    \item \textit{w/o intent parser}: in this case, the model is trained on processed data, but without adopting the intent parser.
    \item \textit{with intent parser}: the data processing pipeline also includes the intent parser. 
\end{itemize}

Table~\ref{tab:automatic_evaluation} shows the results of this analysis. 
The table shows that the data processing aids the Seq2Seq model also without the use of the intent parser, while CodeBERT does not take benefit from the basic data processing steps. The performance of both models significantly increases when the data processing is used in combination with the intent parser. Indeed, the full data processing pipeline improves all the metrics by $\sim31$\% on average for Seq2Seq and by $\sim19$\% on average for CodeBERT when the results of the models are compared without using the data processing process. 
The table also highlights that CodeBERT outperforms the Seq2Seq model across all metrics. We conducted a \textit{paired t-test} and found that the differences between the results obtained by CodeBERT with the intent parser and all the other model configurations are statistically significant for all metrics (at $p<0.05$).

\begin{table}[ht]
\footnotesize
\centering
\caption{Automated evaluation of the translation task. Bolded values are the best performance. IP: Intent Parser. ($*=$ p$<$0.05)}
\label{tab:automatic_evaluation}
\begin{tabular}{
>{\centering\arraybackslash}m{1.5cm}| 
>{\centering\arraybackslash}m{1.25cm}
>{\centering\arraybackslash}m{1.25cm}
>{\centering\arraybackslash}m{1.25cm}|
>{\centering\arraybackslash}m{1.25cm}
>{\centering\arraybackslash}m{1.25cm}
>{\centering\arraybackslash}m{1.25cm}}
\toprule
\multirow{2}{*}{\textbf{Automated}} & 
\multicolumn{3}{c}{\textbf{Seq2Seq}} & \multicolumn{3}{c}{\textbf{CodeBERT}}\\
\textbf{Metrics (\%)} & \textit{w/o data processing} & \textit{w/o IP} & \textit{with IP} & \textit{w/o data processing} & \textit{w/o IP} & \textit{with IP} \\
\midrule
\textit{BLEU-1} & 69.99 & 74.57  & 93.46 & 78.42 & 80.11 & \textbf{94.95*}\\
\textit{BLEU-2} & 64.18 & 69.82 & 91.98 & 75.11 & 75.89  & \textbf{93.61*}\\
\textit{BLEU-3} & 60.09 & 66.35 & 90.87 & 72.75 & 73.15  & \textbf{92.68*}\\
\textit{BLEU-4} & 56.43 & 62.97 & 90.03 & 70.54 & 70.11  & \textbf{91.70*}\\
\textit{ACC} & 39.44 & 51.55 & 82.92 & 69.57 & 67.39 & \textbf{89.75*}\\ 
\bottomrule
\end{tabular}
\end{table}

To estimate the actual goodness of the results, we compared the best performance achieved on the \datasetname{} dataset with the state-of-the-art best performances on the Django dataset \cite{oda2015learning}, a corpus widely used for code generation tasks~\cite{ling2016latent,yin2017syntactic,yin2018tranx,yin2019reranking,hayati2018retrieval,dong2018coarse,gemmell2020relevance,xu2020incorporating} and consisting of $18,805$ pairs of Python statements for the Django Web application framework alongside the corresponding English pseudo-code. 
The state-of-the-art best performances on this dataset provide BLEU-4 score and accuracy equal to $84.70$~\cite{hayati2018retrieval} and $80.20$~\cite{yin2019reranking}, respectively, and are therefore lower than the best results in Table~\ref{tab:automatic_evaluation}.
We attribute these differences to the nature of the assembly language, which is a low-level language. Indeed, even if this work targets the IA-32 processor, which is a CISC architecture, the instruction set of the assembly language is still limited if compared to high-level languages, such as Python, which include a wide number of libraries and functions and, therefore, are more complex to automatically generate.

We also investigate the performance of the code generation task on single-line snippets vs. multi-line snippets by performing a fine-grained evaluation.
Table~\ref{tab:automatic_evaluation_single_vs_multi} shows the performance of CodeBERT (with data processing) for single vs. multi-line snippets. Unsurprisingly, we find that accuracy is negatively affected by the length of snippets, while BLEU scores are higher for multi-line snippets. This is because multi-line snippets are longer, there is more opportunity for BLEU scores to be higher (there can be more n-grams that are matched in longer snippets), in contrast to single line snippets. And likewise, since the accuracy metric is an exact match on the entire snippet, performance on multi-line snippets is lower than for single line snippets.

\begin{table*}[ht]
\centering
\caption{Automatic evaluation of the translation task comparing single-line and multi-line snippets from the test set. Bolded values are the best performance.}
\label{tab:automatic_evaluation_single_vs_multi}
\begin{tabular}{
>{\centering\arraybackslash}m{2.5cm}| 
>{\centering\arraybackslash}m{1.75cm}
>{\centering\arraybackslash}m{1.75cm}}
\toprule
\textbf{Automated  Metrics (\%)} & \textbf{Single-line snippets} & \textbf{Multi-line snippets}\\
\midrule
\textit{BLEU-1} & 93.64 & \textbf{98.14} \\
\textit{BLEU-2} & 92.24 & \textbf{96.86} \\
\textit{BLEU-3} & 91.29 & \textbf{95.84}\\
\textit{BLEU-4} & 90.21 & \textbf{94.91}\\
\textit{ACC} & \textbf{90.51} & 85.42 \\
\bottomrule
\end{tabular}
\end{table*}

This first analysis allows us to conclude that \textit{the state-of-the-art NMT models can be applied for the generation of code used to exploit the software, and provide high performance when used in combination with data processing}.

\subsection{Accuracy of NMT at generating assembly code snippets}
\label{subsec:RQ2}

In \S{}~\ref{subsec:RQ1}, we used the code written by the programmers (i.e., the authors of the shellcodes) as ground truth for the evaluation. Therefore, when the model predicts the assembly code snippets starting from their natural language description, the predicted output is compared to code composing the original shellcode attacks. 
However, since the same English intent can be translated into different but equivalent assembly snippets, automated metrics (such as BLEU scores) are not perfect in that they do not credit semantically correct code that fails to match the reference.
For example, the snippets \texttt{jz label} and \texttt{je label} are semantically identical, even if they use different instructions (\texttt{jz} vs. \texttt{je}). Furthermore, these metrics do not indicate whether the generated code would compile or not. 
Accordingly, we define two new metrics: a generated output snippet (single or multi-line) is considered \textbf{\textit{syntactically correct}} if it is correctly structured in assembly language and compiles correctly.
The output is considered \textbf{\textit{semantically correct}} if the snippet is an appropriate translation in assembly language given the intent description. 
Consider the intent \textit{transfer the contents of the \texttt{ebx} register into the \texttt{eax} register}. If the approach generates the snippet \texttt{mov ebx, eax}, then the snippet is considered syntactically correct (it would compile), but not semantically correct because the order of the operands is inverted. 
These two metrics allow us to assess the deeper linguistic features of the code~\cite{han-etal-2021-translation}. The semantic correctness implies syntax correctness, while a snippet can be syntactically correct but semantically incorrect. When a snippet is syntactically incorrect it is also semantically incorrect. The evaluation of the semantic equivalence between the output predicted by the models and the code written by the authors of the shellcodes provides the best insights into the quality of the output since it allows us to assess the correctness of the predicted code even if its syntax differs from the ground truth. This is the reason why we did not limit the analysis to automatic metrics, and manually evaluated the semantic meaning of generated code.

To evaluate the syntactic correctness of the outputs, we used the NASM compiler in order to check whether the code is compilable, while we evaluated the semantic correctness by checking if the code generated by the models is a correct translation of the English intent. 
We performed this analysis manually, by checking every single line of generated code. This analysis could not be performed automatically, since an English intent can be translated into several forms that are different, but semantically equivalent. For the same reason, manual (“human”) evaluation is a common practice in NMT studies. The manual evaluation also gives better insights into the quality of machine translation and allows us to analyze errors in the output. To reduce the possibility of errors in manual analysis, multiple authors performed this evaluation independently, obtaining a consensus for the semantic correctness of the output predicted by the models.

Table~\ref{tab:manual_evaluation} shows the percentage of syntactically and semantically correct snippets across all the examples of the test set. We evaluated the performance of Seq2Seq and CodeBERT, both using data processing. Both syntactic and semantic evaluations were performed by compiling the generated snippets under the NASM compiler. 
Table~\ref{tab:manual_evaluation} shows that both approaches are able to generate  $>95\%$ of syntactically correct snippets. 
\textit{Paired t-tests} indicated that the differences between the models are not statistically significant for the syntactic correctness, but they are statistically significant for semantic correctness (at $p<0.01$).

\begin{table}[ht]
\centering
\caption{Code correctness evaluation of the translation task given the whole test set. Bolded values are the best performance. ($*=$ p$<$0.01)}
\label{tab:manual_evaluation}
\begin{tabular}{
>{\centering\arraybackslash}m{3cm}| >{\centering\arraybackslash}m{1.5cm}
>{\centering\arraybackslash}m{1.5cm}}
\toprule
\textbf{Code Correctness Metrics  (\%)}  & \textbf{Seq2Seq with data processing} & \textbf{CodeBERT with data processing}\\
\midrule
\textit{Syntactically Correct} & 96.58 & \textbf{97.20} \\
\textit{Semantically Correct} & 85.40 & \textbf{93.16*} \\
\bottomrule
\end{tabular}
\end{table}

Again, we further investigated the results provided by CodeBERT, by evaluating the performance of the model on single vs. multi-line snippets. 
Table~\ref{tab:manual_evaluation_single_vs_multi} highlights that the multi-line snippets affect model performance on syntactic correctness, although we find no statistically significant difference in model performance on the semantic correctness metric.

\begin{table}[ht]
\centering
\caption{Code correctness evaluation of the translation task comparing single-line and multi-line snippets from the test set. Bolded values are the best performance.}
\label{tab:manual_evaluation_single_vs_multi}
\begin{tabular}{
>{\centering\arraybackslash}m{3cm}| >{\centering\arraybackslash}m{1.6cm}
>{\centering\arraybackslash}m{1.6cm}}
\toprule
\textbf{Code Correctness Metrics (\%)} & \textbf{Single-line snippets} & \textbf{Multi-line snippets}\\
\midrule
\textit{Syntactically Correct} & \textbf{97.81} & 93.75 \\
\textit{Semantically Correct} & 93.06 & \textbf{93.75} \\
\bottomrule
\end{tabular}
\end{table}

Table~\ref{tab:successful_cases} show illustrative examples of code snippets that the model can successfully translate (i.e., the snippets generated by the approach are syntactically and semantically correct). 
Rows 3, 6, and 8 are examples of correct snippets that are penalized by automated metrics, even if they do not exactly match the ground truth. 
Despite some slight differences with the ground truth, the generated code is semantically correct, due to the ambiguity of the assembly language. Thus, these differences are still considered correct by our manual analysis. 
We note correctly generated examples of multi-line snippets in rows 2, 3, 4, and 6. Also, we observe in row 3, the ability to generate multi-line snippets from a relatively abstract intent.

We conclude that \textit{both Seq2Seq and CodeBERT provide syntactically and semantically correct code snippets with high accuracy. Moreover, CodeBERT provides the best performance in the task of generating shellcodes from natural language intents.}
Due to these findings, we consider CodeBERT (with data processing) as our reference NMT model for the following experimental objectives.

\begin{table*}[t]
\centering
\caption{Illustrative examples of successfully generated snippets using our approach. Differences between the output and ground truth are bolded. Such differences are penalized by automatic metrics even though they are correct.}
\label{tab:successful_cases}
\begin{tabular}{
>{\centering\arraybackslash}m{0.75cm}|
 >{\centering\arraybackslash}m{3.5cm}|
 >{\centering\arraybackslash}m{3cm}|
 >{\centering\arraybackslash}m{3cm}}
\toprule
\textbf{Row} & \textbf{Natural Language Intent} & \textbf{Ground Truth} & \textbf{Model Output}\\
\midrule
1 & \textit{Move the byte at the address [edi] into bl} & \texttt{mov bl, byte [edi]} & \texttt{mov bl, byte [edi]} \\
\midrule
2 & \textit{Jump to the \_start label if the value in the eax register is not equal to the doubleword addressed by edi} & \texttt{scasd \textbackslash{n} jnz \_start} &	\texttt{scasd \textbackslash{n} jnz \_start} \\
\midrule
3 & \textit{Put /bin/sh into ebx} & \texttt{push 0x68732f2f \textbackslash{n} push 0x6e69622f \textbackslash{n} mov ebx, esp} &\texttt{push \textbf{long} 0x68732f2f \textbackslash{n} push \textbf{long} 0x6e69622f \textbackslash{n} mov ebx, esp} \\
\midrule
4 & \textit{Push the value 0x61702f2f and the value 0x6374652f onto the stack and point the ebx register to the stack register} & \texttt{push 0x61702f2f \textbackslash{n} push 0x6374652f \textbackslash{n} mov ebx, esp} & \texttt{push 0x61702f2f \textbackslash{n} push 0x6374652f \textbackslash{n} mov ebx, esp} \\
\midrule
5 & \textit{Perform a logical xor between the address specified by [ecx + 116] and the dh register and save the result in [ecx + 116]} & \texttt{xor [ecx + 116], dh} & \texttt{xor [ecx + 116], dh} \\
\midrule
6 & \textit{If the contents of the al register is equal to the value 0xf2 then jump to the \_start label} & \texttt{cmp al, 0xf2 \textbackslash{n} jz \_start} & \texttt{cmp al, 0xf2 \textbackslash{n} \textbf{je} \_start} \\
\midrule
7 & \textit{Move esi into the doubleword starting at the address esp-4} & \texttt{mov dword [esp-4], esi} & \texttt{mov dword [esp-4], esi} \\
\midrule
8 & \textit{Call kernel}	& \texttt{int 0x80} & \texttt{int 0x80\textbf{h}} \\
\bottomrule
\end{tabular}
\end{table*}

\subsection{Accuracy of the NMT at generating whole shellcodes}
\label{subsec:RQ3}
The ultimate goal of developers is to craft entire shellcodes. The previous evaluation showed that NMT can generate individual code snippets that are likely the correct ones. Thus, NMT can be queried by developers to translate specific parts of a program. Here, we raise the bar for the evaluation, by analyzing to which extent NMT can generate an \emph{entire} shellcode. To this purpose, we consider groups of intents from the same exploit and compare the resulting code snippets with the original shellcode. 
We use two new metrics to evaluate the ability of the approach to generate semantically and syntactically correct code for entire shellcodes.

Let $n_t^i$ be the the number of total lines of the $i$-th program in the test set ($i \in [1,30]$). Let also consider $n_{syn}^i$ as the number of automatically-generated snippets for the $i$-th program that are syntactically correct, and $n_{sem}^i$ as the number of automatically-generated snippets that are semantically correct. For every program of the test set, we define the \textit{\textbf{syntactic correctness}} of the program $i$ as the ratio $n_{syn}^i / n_t^i$, and the \textit{\textbf{semantic correctness}} of the program as the ratio $n_{sem}^i / n_t^i$. 
To perform a conservative evaluation on multi-line snippets, even if only one line of code of the generated snippets is syntactically (semantically) incorrect, we consider all the lines belonging to the multi-line block as syntactically (semantically) incorrect. 
Both metrics range between $0$ and $1$.

For each $i \in [1,30]$, we computed the values $n_{syn}^i$ and $n_{sem}^i$ for the assembly programs in the test set. 
We found that the average syntactic correctness over all the programs of the test set is $\sim 98\%$ (standard deviation is $\sim 4\%$). Similarly, we estimated the average semantic correctness, which is equal to $\sim96\%$ (standard deviation is $\sim 6\%$). Out of $30$ programs, we found that $21$ are \textbf{\textit{compilable}} with NASM and \textbf{\textit{executable}} on the target system.



\begin{figure}[ht]

    \centering
    
    \subfloat[Density plot.\label{fig:density_plot}]{%
    \includegraphics[width=0.49\textwidth]{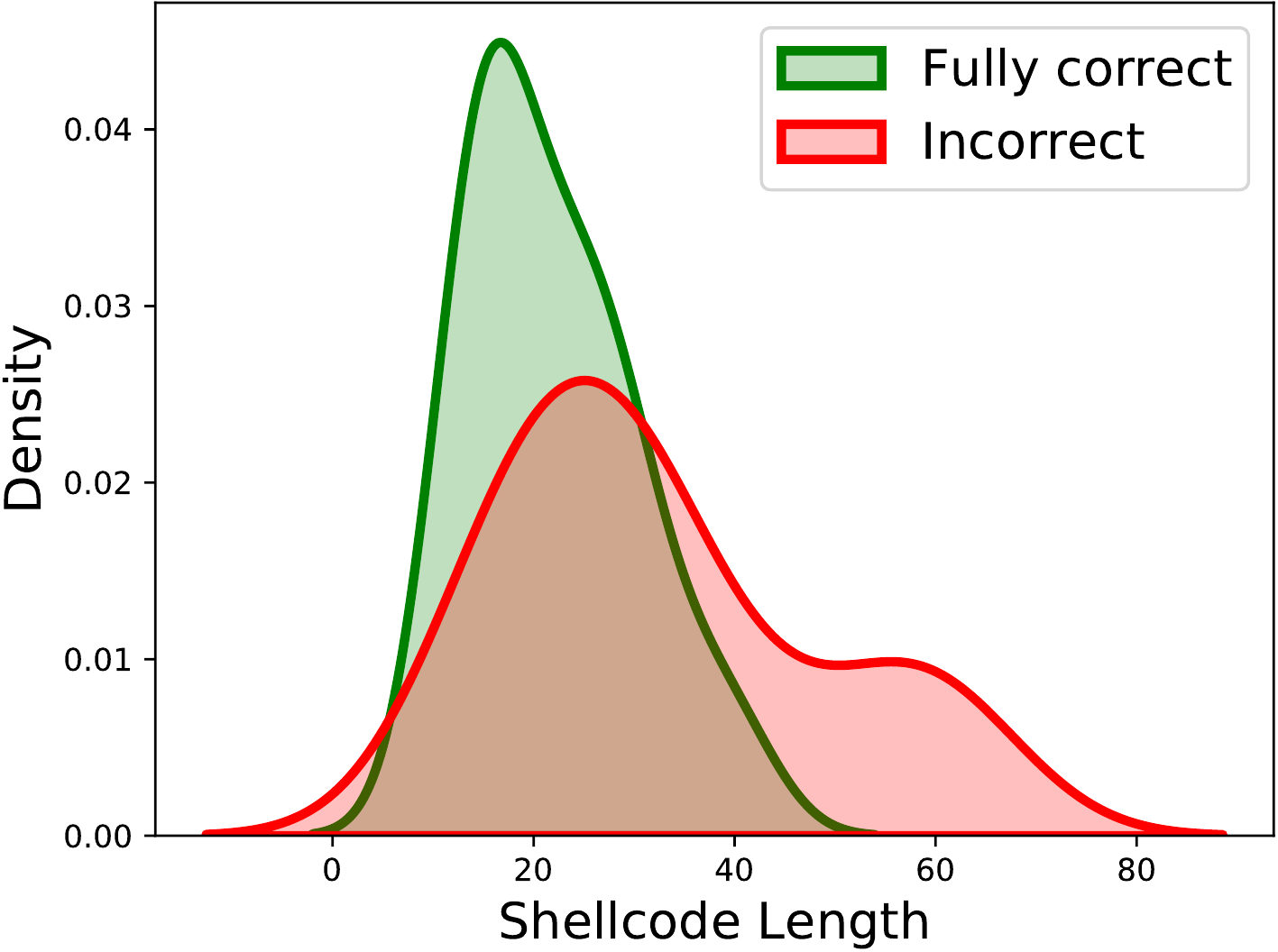}}
    \hfill
    \subfloat[Box plot.\label{fig:time_events}]{%
    \includegraphics[width=0.49\textwidth]{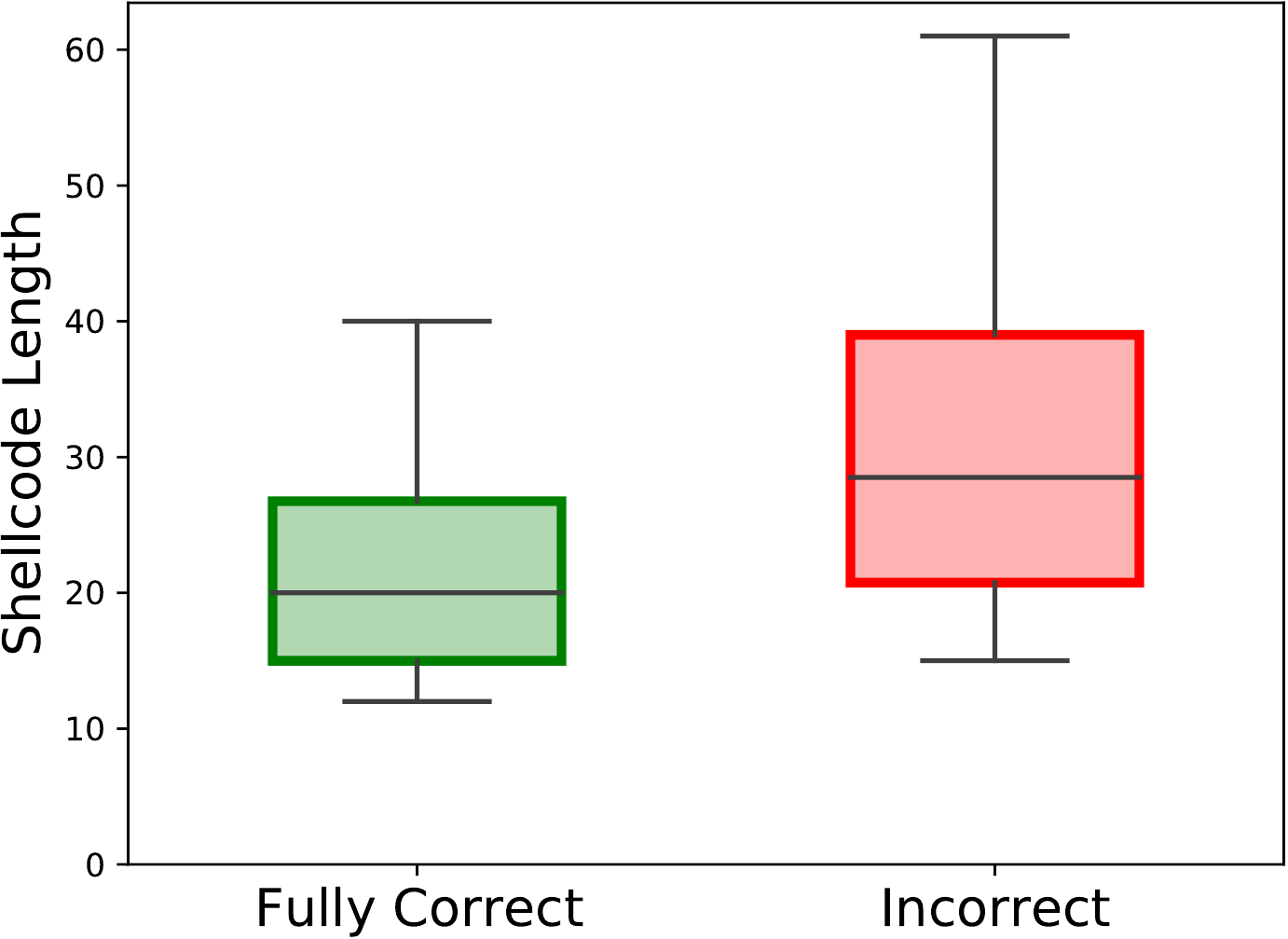}}
    \caption{Plots visualizing the statistics, in terms of lines of assembly code, of the  30 shellcodes in the test set. The labels \textit{Fully Correct} and \textit{Incorrect} refer to the shellcodes that are generated by the approach as fully correct ($n_{sem}/n_t = 1$) and incorrect ($n_{sem}/n_t < 1$), respectively..}
    \label{fig:shellcode_length}
     
\end{figure}

Since even one incorrect line of code suffices to thwart the effectiveness of a shellcode, we analyzed how many shellcodes could be generated with no errors. We consider a shellcode as \textit{\textbf{fully correct}} if all the assembly instructions composing the shellcode are individually semantically correct (i.e., $n_{sem}^i/n_{t}^i = 1$).
This evaluation metric is a demanding one. Even if one single line of the shellcode is not semantically correct, then the whole program is considered as not correctly generated. Despite this conservative evaluation, our approach is able to correctly generate 16 out of 30 whole shellcodes. 
Figure~\ref{fig:shellcode_length} shows the summary statistics with a density and a box plot, differentiating the \textit{fully correct} shellcodes  from the \textit{incorrect} ones. As expected, the complexity of the shellcode - in terms of lines of assembly code - impacts the ability of the approach to correctly generate the whole program. 
However, the average (and the median) length of the shellcodes incorrectly generated by the model is affected by the three assembly programs of lengths 55, 59, and 61. If we consider these shellcodes as outliers, then the group of fully correct shellcodes and the group of the incorrectly generated shellcodes are very similar in terms of size.
\textit{We interpret these results as a promising indication towards our ultimate goal of generating entire shellcode programs automatically from short natural language intents.}

\subsection{Types of errors incurred by NMT in the generation of shellcodes}
\label{subsec:RQ4}

In the last experiment objective, we performed a manual inspection of the model's mispredictions. We noticed that the failure outputs fall down in the following three failure types;
\begin{itemize}
    \item \textbf{\textit{Failure Type A}}: translation failure in generating the correct label, instruction, operand(s), or delimiter(s).
    \item \textbf{\textit{Failure Type B}}:  translation failure in identifying the correct order and/or the addressing mode of operands.
    \item \textbf{\textit{Failure Type C}}: intent parser's failure in identifying one or more of the explicitly stated identifiers.
\end{itemize}

The failure types A and B are due to the lack of ability of the model to perform the correct translation of the English intent in the assembly code. The failure type C, instead, is attributed to the intent parser failure. Indeed, even if the performance of the translation task benefits from the work of the intent parser (see \S{}~\ref{subsec:RQ1}), it is not perfect and can lead to a failure prediction by wrongly identifying the variable or register names, labels, etc. 

Moreover, the error predictions can be further classified as syntactically incorrect and semantically incorrect. We remark that the syntactic incorrectness implies the semantic one.
To better illustrate the problem, we present in Table~\ref{tab:failure_cases} a qualitative evaluation using cherry- and lemon-picked examples of failure prediction from our test set.

\begin{table*}[ht]
\scriptsize
\centering
\caption{Illustrative examples of incorrect outputs. The prediction errors are \textcolor{red}{\textbf{red/bold}}. \textcolor{red}{\textbf{\cancel{Slashed}}} text refers to omitted predictions. \textit{\textbf{Syn}} indicates a syntactically and semantically incorrect snippet, while \textit{\textbf{Sem}} indicates a semantically incorrectness output.} 
\label{tab:failure_cases}
\begin{tabular}{
>{\centering\arraybackslash}m{0.5cm}|
 >{\centering\arraybackslash}m{3cm}|
 >{\centering\arraybackslash}m{2.75cm}|
 >{\centering\arraybackslash}m{2.75cm}|
 >{\centering\arraybackslash}m{1cm}}
\toprule
\textbf{Row} &
\textbf{Natural Language Intent} & \textbf{Ground Truth} & \textbf{Model Output} & {\textbf{Failure Type}}\\
\midrule
1 & \textit{Perform the xor operation between the location pointed by ecx and dh} & \texttt{xor [ecx], dh} & \texttt{xor \textcolor{red}{\textbf{ecx, [dh]}}} & {\textbf{\textit{B, Syn}}}\\
\midrule
2 & \textit{Jump to the \_start label if the value in the eax register is not equal to the doubleword addressed by edi else jump to the edi register} & \texttt{scasd \textbackslash{n} jnz \_start \textbackslash{n} jmp edi} & \texttt{scasd \textcolor{red}{\textbf{\textbackslash{\textbackslash}}} jnz \_start \textbackslash{n} jmp edi} & {\textbf{\textit{A, Syn}}} \\
\midrule
3 & \textit{Define the array of bytes encodedshellcode and initialize it to 0x32, 0x51, 0x30, 0x74, 0x69, 0x63, 0x6f, 0xe4, 0x8a, 0x54, 0xe2, 0x0c, 0x81, 0xc1, 0x69, 0x30, 0x69, 0x30, 0x6a, 0x8a, 0x51, 0xe3, 0x8a, 0xb1, 0xce} & \texttt{encodedshellcode: db 0x32, 0x51,0x30, 0x74, 0x69, 0x63, 0x6f, 0xe4, 0x8a, 0x54, 0xe2, 0x0c, 0x81, 0xc1, 0x69, 0x30, 0x69, 0x30, 0x6a, 0x8a, 0x51, 0xe3, 0x8a, 0xb1, 0xce} & \texttt{\textcolor{red}{\cancel{\textbf{encodedshellcode:}}} db 0x32, 0x51, 0x30, 0x74, 0x69, 0x63, 0x6f, 0xe4, 0x8a, 0x54, 0xe2, 0x0c, 0x81, 0xc1, 0x69, 0x30, 0x69, 0x30, 0x6a, 0x8a, 0x51, 0xe3, 0x8a, 0xb1, 0xce} &{\textbf{\textit{A, Sem}}} \\
\midrule
4 & \textit{Call decoder} &    \texttt{call decoder} & \texttt{\textcolor{red}{\cancel{\textbf{call}}} decoder\textcolor{red}{\textbf{:}}}& {\textbf{\textit{A, Sem}}} \\
\midrule
5 & \textit{Decrement the ecx register and jump to the main\_push label if the contents of the ecx register is not zero else move the value 30 into the cl register} & \texttt{loop main\_push \textbackslash{n} mov cl, 30} & \texttt{loop main\_push \textbackslash{n} mov \textcolor{red}{\textbf{var3}}, \textcolor{red}{\textbf{esp}}} & {\textbf{\textit{C, Syn}}} \\
\midrule
6 & \textit{Perform logical or between the cx register and 0xfff} &    \texttt{or cx, 0xfff} & \texttt{\textcolor{red}{\textbf{x}}or \textcolor{red}{\textbf{0xfff}}, \textcolor{red}{\textbf{cx}}}& {\textbf{\textit{A, B, Syn}}} \\
\midrule
7 & \textit{Perform a logical xor between the location pointed by [ecx + 116] and the bh register and save the result in [ecx + 116]} &    \texttt{xor [ecx + 116], bh} & \texttt{xor byte [ecx + 116], \textcolor{red}{\textbf{0x3}}}& {\textbf{\textit{A, C, Sem}}} \\
\midrule
8 & \textit{Move 2001Q into cx} & \texttt{mov cx, 2001Q} & \texttt{mov cx, 2001\textcolor{red}{{\cancel{\textbf{Q}}}}}& {\textbf{\textit{C, Sem}}} \\
\midrule
9 & \textit{Push the word 0666q onto the stack} & \texttt{push word 0666q} & \texttt{push word 06\textcolor{red}{\textbf{44o}}}& {\textbf{\textit{C, Sem}}} \\
\midrule
10 & \textit{Declare the read label} & \texttt{read:} & \texttt{\textcolor{red}{\textbf{section}}:}& {\textbf{\textit{A, C, Syn}}}\\
\bottomrule
\end{tabular}
\end{table*}

The first row showcases an example of failing to model because of implicit knowledge. The intent does not mention the indirect addressing mode (specified by the bracket [] in NASM syntax). 
In the second row, we note that the model failed to generate the newline token properly to separate the snippets with lines. This causes a syntax issue, and since it does not compile we count it as syntactically incorrect.
The third row shows an example in which a byte string is declared without defining the label, while the fourth row illustrates the model's failure to predict the right instruction (the definition of the function \texttt{decoder} instead of the execution of the function). Both outputs do not raise an exception when compiled, therefore they are syntactically but not semantically correct.
In the fifth row, we note that the intent parser correctly identifies \texttt{main\_push} in the standardization process, but fails to recognize the \texttt{cl} register and misidentifies \texttt{ecx} instead. We also note that the model predicted a \texttt{mov} operation between two registers (\texttt{register, esp}) rather than a register and a value.  The predicted register does not exist in the intent hence, the output is a \texttt{var3}. 
The sixth row shows an example with incorrect instruction and inverse operands order.
The remaining examples include the intent parser failing to identify explicitly stated identifiers or letters in values sometimes in long intents such as in the case of the \texttt{bh} register (row 7) and occasionally in simple contexts such as in the case of \texttt{read} (row 10). The last row is considered also syntactically incorrect since it is not possible to declare a label with the \texttt{section} assembly directive. 
This goes to show when there is a mistake in the standardization step, the translation may fail to work around it even if the intent seems simple.

The failure outputs also provide indications on what it can be done to increase the performance of the code generation task. Most of the errors can be easily identified by the programmers: incorrect addressing modes (first row),  wrong newline character (second row), missing labels (e.g., \texttt{encodedshellcode} in row number 3), wrong instructions (row 4, 6), undefined variables (e.g., \texttt{var3} in row 5), wrong operand orders (row number 6), etc.
The syntactically incorrect predictions, i.e., the predictions that do not follow the syntax, can be identified with a compiler and can be fixed through an ``intelligent" post-processing phase, which should be trained to identify and fix the failure outputs. This is part of the future work.

\subsection{Discussion and Lessons Learned}
\label{subsec:discussion}
The experimental analysis pointed out that NMT models can efficiently generate assembly code for real shellcodes, starting from their natural description. When used in combination with data processing, the accuracy of the code generation task is high enough to support developers in developing software exploits. 
Even if the size and the complexity of an English intent increase, the performance of the translation task is not negatively affected. 
CodeBERT achieves the best performance and further justifies its wide usage to address software engineering tasks. 
The model is able to generate whole software exploits with syntactic and semantic correctness greater than $95\%$. It is also able to generate programs that are fully correct, i.e., compilable and executable on the target system. However, the complexity of the software attacks (in terms of lines of code) reduces the accuracy of generating entire programs.
The analysis also pointed out that the most common error predictions are easily identifiable and can be fixed during the post-processing process.


\section{Ethical Considerations}
\label{sec:ethics}

Recognizing that attackers use exploit code as a weapon, it is important to specify that the goal of the \textit{proof-of-concept} (POC) exploits is not to cause harm but to surface security weaknesses within the software. Identifying such security issues allows companies to patch vulnerabilities and protect themselves against attacks. 

\emph{Offensive security} is a sub-field of security research that tests security measures from an adversary or competitor’s perspective. It can employ ethical hackers to probe a system for vulnerabilities \cite{bugcrowd,hackerone,github_blog}. \emph{Automatic exploit generation} (AEG), an offensive security technique, is a developing area of research that aims to automate the exploit generation process and to explore and test critical vulnerabilities before they are discovered by attackers~\cite{aeg}. Indeed, work such as ours, which studies exploits on compromised systems can provide valuable information about the technical skills, degree of experience, and intent of the attackers. By using this information, it is possible to implement measures to detect and prevent attacks \cite{arce2004shellcode}.



\section{Threats to Validity}
\label{sec:threats}
\noindent
\textbf{NMT models:} Before the era of NMT, Statistical Machine Translation (SMT)~\cite{costa2014statistical} was the most popular technique for software engineering (SE) problems, it still outperforms NMT in some SE problems~\cite{phan2020statistical}. However, since we are interested in the specific problem of code generation, we focus on NMT that has shown superior performance on public benchmarks~\cite{bojar2016findings}, and that it is widely recognized as the premier method for the translation of different languages~\cite{wu2016google}. 
Our choice of the NMT models has been influenced by their popularity and the availability of mature open-source implementations. We acknowledge that using only two state-of-the-art models can be a limitation of this work. Nevertheless, we believe that these two models are valid representatives of the NMT research area, and can provide us with a realistic evaluation of NMT for code generation. 
Seq2Seq has been for several years the most used model for code generation tasks, and it is still widely employed in NMT studies as a baseline model. 
CodeBERT has pushed the boundaries in natural language processing and represents the state-of-the-art for generating code documentation given snippets, as well as retrieving code snippets given a natural language search query across six different programming languages~\cite{husain_codesearchnet_2019}. Moreover, it has also been applied in software engineering to perform different tasks \cite{pan2021empirical}.

\vspace{0.1cm}
\noindent
\textbf{Size of our dataset:} Our dataset contains $3,200$ instances, which may seem relatively small compared to training data available for other NLP tasks. 
The data about shellcodes is much more difficult to obtain than other data for NMT. For example, before starting the collection of the dataset, we developed a script to collect assembly code for IA-32 from all of the repositories on GitHub (by far the source most used by empirical software engineering studies). We found that the amount of available data is very limited. The data is further restricted by the fact that we are specifically interested in security-oriented assembly codes (i.e., shellcodes). Therefore, we decided to collect all the shellcodes for Linux/IA-32 from exploit-db and shell-storm, the two public databases for shellcodes most popular among the security professionals, to achieve representativeness. We collected shellcodes written over a large period (from 2000 to 2020) from a variety of authors, in order to achieve diversity. To the best of our knowledge, the resulting dataset is the largest collection of shellcodes in assembly available to date.
Despite the previous considerations, we note that our dataset is comparable in size to the popular CoNaLa dataset~\cite{yin2017syntactic} ($2,379$ training and $500$ test samples in the \textit{annotated} version of the dataset), which is the basis for state-of-the-art studies in NMT for Python code generation \cite{yin2018mining,yin2019reranking,gemmell2020relevance}. Further, \datasetname{} contains a higher percentage of multi-line snippets ($\sim16\%$ vs. $\sim4\%$). We also note here that existing code generation datasets do contain a larger, potentially noisy, subset of training examples (ranging in several thousand) obtained by mining the web. For example, the CoNaLa \textit{mined} (as opposed to the CoNaLa \textit{annotated}) dataset contains $598,237$ training examples mined directly from  StackOverflow~\cite{yin2018mining}. 
We designed the proposed approach to leverage existing pre-trained models to compensate for the need for big data, by training the model using our assembly dataset.

\vspace{0.1cm}
\noindent
\textbf{Code description:} To build the dataset, we described in the English language the shellcodes collected from publicly available exploit databases. Therefore, the description of the assembly code derives from our considerations and knowledge.
However, the building process of the \datasetname{} dataset is not different from other corpus built from scratch. For example, Oda \textit{et al.} \cite{oda2015learning} hired an engineer to create pseudo-code for the Django Web application framework and obtain the corpus. 
We avoided a single centralized version of the code description to take into account the variability of descriptions in natural language. Indeed, multiple authors described independently different samples of the dataset in the English language, and, where available, we kept untouched the comments written by developers of the collected programs to describe the assembly code snippets.
To understand how different programmers and experts describe the assembly code for IA-32 and how to deal with the ambiguity of natural language in this specific context, we took inspiration from popular tutorials and books \cite{duntemann2011assembly,kusswurm2014modern,tutorialspoint}.

\vspace{0.1cm}
\noindent
\textbf{Translation task:} As assembly code is a low-level language, it often takes a long sequence of instructions to complete an atomic function. Therefore, some translations presented in the dataset are too ``literal'' and cumbersome. For example, instead of writing \textit{``Define the \_start label''}, a user might just as well write ``\texttt{\_start:}'', similarly, the intent \textit{``Push the contents of eax onto the stack''} takes longer than writing the assembly instruction ``\texttt{push eax}''. 
However, this is a common situation in any translation task from English to programming language. For example, the Django dataset 
contains numerous Python code snippets that are relatively short (e.g., ``\texttt{chunk\_buffer = BytesIO(chunk)}'') described with with English statements that are definitely longer than the snippets (\textit{``evaluate the function BytesIO with argument chunk, substitute it for chunk\_buffer.''}). Similarly, in the CoNaLa dataset we can find shortcode snippets (e.g., ``\texttt{GRAVITY = 9.8}'') described with longer English intents (\textit{``assign float 9.8 to variable GRAVITY''}). 
Nevertheless, we -- and other datasets-- still include such verbose intents to provide richer learning of NMT models. Moreover, we mitigated this problem by adding multi-line snippets, i.e., single intents described in natural language that generate more lines of assembly codes, that are closer to the intent that developers may want to use during development.


\vspace{0.1cm}
\noindent
\textbf{Scope of the approach:} A shellcode is a piece of assembly code written specifically for exploitation purposes. From this perspective, all shellcodes are security-related programs and, therefore, the proposed approach is tailored for generating software exploits. It is an interesting question whether the proposed approach has applications beyond security. 
The approach is focused on assembly programs, which is the most used language for shellcodes. Thus, the processing pipeline has been designed to handle relevant elements of the assembly language, such as keywords and register names. This approach significantly contributes to generating more accurate code compared to generic NMT techniques but narrows the scope to assembly code. As future work, we are exploring the use of NMT for other programming languages, such as Python.
In principle, a programmer can use the method to generate assembly code unrelated to security applications. However, the method might be less accurate in this case, since our solution is trained with a dataset of mostly security-related assembly code snippets. To be used outside security applications, the programmer would need to adopt a training dataset with more non-security assembly code (e.g., assembly code for device drivers or microcontrollers). Moreover, it may be necessary to tweak the processing pipeline to support special keywords that are not adopted for shellcodes (e.g., linking directives for embedded software). We opted to leave such extensions out of the scope of our work, as security applications are the ones that have by far the highest demand for increasing the productivity of assembly programming.

\section{Conclusion and Future Work}
\label{sec:conclusion}
We addressed the problem of automated exploit generation using natural language processing techniques. We use Neural Machine Translation to translate natural language intents into shellcode. We built and released the first dataset of shellcodes, \textit{Shellcode\_IA32}, containing $3,200$ pairs of code snippets and intents. The dataset also contains $510$ intents that generate multiple snippets.
These assembly language snippets can be combined to generate shellcodes for the Intel 32-bit Architecture. Our empirical analysis demonstrated the feasibility of using NMT for this task, using both automated and manual metrics. 
We also propose the use of novel metrics for the task of code generation, that we anticipate would be useful to the community. 


Our work enables further studies in the area, to make NMT more and more effective. 
We are currently working on a new engine for the post-processing phase, in order to identify and fix the assembly lines wrongly generated by the NMT model and to further improve accuracy. 
We are also analyzing the impact of ``noisy inputs'' or ``perturbation'' in the natural language, since human developers may provide inaccurate or incomplete descriptions of the shellcode to be generated. For example, perturbations can be introduced by replacing words with ``unseen'' synonyms, or by removing redundant information. In this direction, we are investigating a solution to make NMT more robust and usable, by helping the model to derive the missing information (i.e., information not explicitly stated in the English intent) from the context of the programs.
Finally, as part of future research, we aim to evaluate our approach with actual humans instructing with comments, so that the evaluation could take into account how the humans perceive the actual usefulness of developing a shellcode that achieves the desired result.

Beyond our current work on extending the proposed approach, we expect that this work can support more researchers in the field. 
Indeed, in the era where deep learning is evolving at a quick pace and succeeding in more and more tasks with surprising accuracy, we expect in the near future the development of new deep learning architectures, which could potentially bring benefits for the automatic generation of exploits. In this light, the proposed approach and dataset represent valid means to pave the way for a new generation of offensive security methods. 
This work represents a first step towards the ambitious goal of automatically generating shellcodes from natural language, provides originally-collected data, enables replication, and describes successes and challenges through rigorous evaluation. 



\appendix
\label{appendix}
\section{Test Set}
\label{appendix:test_set}
Table~\ref{tab:shellcodes_stats} presents detailed information on the $30$ shellcodes composing the test set. In particular, the table shows the URL where the shellcode is collected, the number of assembly lines of the program, the number of multi-line snippets, and the number of snippets generated incorrectly from our approach. We consider the whole shellcode generated correctly only if the approach produces $0$ incorrect snippets. Our approach generated correctly $16$ out of $30$ whole shellcodes.

\begin{table*}[ht]
\centering
\caption{The 30 shellcodes composing the test set. We consider a shellcode executed correctly if all the generated snippets composing the program are semantically correct.
$n_t$: number of total assembly lines of the program.
\textit{Multi-line}: number of multi-lines snippets in the program.
$n_{syn}$: number of syntactically correct lines generated by the approach. 
$n_{sem}$: number of semantically correct lines generated by the approach.}
\label{tab:shellcodes_stats}
\scriptsize
\begin{tabular}
{>{\centering\arraybackslash}m{0.25cm}| 
>{\centering\arraybackslash}m{6cm}
>{\centering\arraybackslash}m{1.5cm}
>{\centering\arraybackslash}m{1cm}
>{\centering\arraybackslash}m{1cm}}
\toprule
\textbf{id} & \textbf{URL} & $n_{t}$ \textit{(Multi-line)} & $n_{syn}$ & $n_{sem}$\\ \midrule
1 & \url{www.exploit-db.com/shellcodes/13452} & 17 (4) & 15 & 12\\
  2 & \url{www.exploit-db.com/shellcodes/48703} & 33 (16) & 31 & 29\\
 3 & \url{www.exploit-db.com/shellcodes/47877} & 40 (0) & 40 & 40\\
 4 & \url{www.exploit-db.com/shellcodes/13716} & 59 (0) & 58 & 52\\
 5 & \url{www.exploit-db.com/shellcodes/47513} & 14 (0) & 14 & 14\\
 6 & \url{www.exploit-db.com/shellcodes/47511} & 24 (0) & 24 & 24\\
 7 & \url{www.exploit-db.com/shellcodes/47481} & 41 (2) & 40 & 38\\
 8 & \url{www.exploit-db.com/shellcodes/47396} & 61 (15) & 61 & 60\\
 9 & \url{www.exploit-db.com/shellcodes/47200} & 29 (2) & 29 & 28\\
 10 & \url{www.exploit-db.com/shellcodes/47202} & 29 (4) & 29 & 29\\
 11 & \url{www.exploit-db.com/shellcodes/47108} & 26 (9) & 26 & 26\\
 12 & \url{www.exploit-db.com/shellcodes/47068} & 12 (0) & 12 & 12\\
 13 & \url{www.exploit-db.com/shellcodes/46994} & 28 (4) & 27 & 26\\
 14 & \url{www.exploit-db.com/shellcodes/46829} & 20 (6) & 20 & 20\\
 15 & \url{www.exploit-db.com/shellcodes/46801} & 34 (9) & 34 & 34\\
 16 & \url{www.exploit-db.com/shellcodes/46791} & 27 (8) & 27 & 26\\
 17 & \url{www.exploit-db.com/shellcodes/46704} & 29 (6) & 29 & 29\\
 18 & \url{www.exploit-db.com/shellcodes/46704} & 55 (4) & 55 & 54\\
 19 & \url{www.exploit-db.com/shellcodes/45669} & 20 (6) & 20 & 20\\
 20 & \url{www.exploit-db.com/shellcodes/45940} & 25 (4) & 25 & 25\\
 21 & \url{www.exploit-db.com/shellcodes/45529} & 14 (7) & 14 & 14\\
 22 & \url{www.exploit-db.com/shellcodes/45441} & 20 (9) & 17 & 17\\
 23 & \url{www.exploit-db.com/shellcodes/44963} & 17 (6) & 17 & 17\\
 24 & \url{www.exploit-db.com/shellcodes/44609} & 32 (0) & 31 & 30\\
 25 & \url{www.exploit-db.com/shellcodes/44509} & 16 (2) & 16 & 16\\
 26 & \url{www.exploit-db.com/shellcodes/44594} & 15 (2) & 15 & 15\\
 27 & \url{www.exploit-db.com/shellcodes/44510} & 23 (3) & 23 & 21\\
 28 & \url{www.exploit-db.com/shellcodes/43476} & 15 (6) & 15 & 15\\
 29 & \url{www.exploit-db.com/shellcodes/43489} & 18 (2) & 17 & 17\\
 30 & \url{www.exploit-db.com/shellcodes/43463} & 15 (3) & 15 & 14\\ \bottomrule
\end{tabular}
\end{table*}



\newpage

\begin{acknowledgements}
This work has been partially supported by the University of Naples Federico II in the frame of the Programme F.R.A., project id OSTAGE. 
\end{acknowledgements}

%
%

\bibliographystyle{spmpsci}      
\bibliography{bibliography}   

%
%

\end{document}